\documentclass{article}

\usepackage{amsmath}
\usepackage{booktabs}
\usepackage{color}
\usepackage{listings}
\usepackage{subcaption}
\usepackage{url}
\usepackage{tikz}
\usepackage{gnuplot-lua-tikz}
\usepackage{adjustbox}

\renewcommand\t[1]{{\tt #1}}

\newcommand\lt[1]{{\lstinline+#1+}}
\definecolor{dkgreen}{rgb}{0,0.5,0}
\definecolor{dkred}{rgb}{0.5,0,0}
\definecolor{gray}{rgb}{0.5,0.5,0.5}
\usetikzlibrary{trees}

\def\speca{(0,0)   circle (2.5cm)}
\def\specb{(0,0)   circle (3cm)}
\def\specc{(0,0)   circle (3.5cm)}
\def\test{(0:2cm) circle (3.5cm)}
\def\equiv{(-1cm:1.5cm) circle (1cm)}
\def\orig{(1.25,-0.5) circle (0.25cm)}
\def\muta{(1.5,1.5) circle (0.25cm)}
\def\mutb{(1.25,-1.25) circle (0.25cm)}
\def\mutc{(-2,-0.5) circle (0.25cm)}
\def\mutd{(4,1) circle (0.25cm)}
\def\mute{(-3,3) circle (0.25cm)}

\tikzstyle{label} = [fill opacity=1, font=\large, text width=8em]
\tikzstyle{class} = [fill opacity=1]
\tikzstyle{gre} = [circle, draw, fill=green!80]
\tikzstyle{blu} = [circle, draw, fill=blue!80]
\tikzstyle{yel} = [circle, draw, fill=yellow]
\tikzstyle{red} = [circle, draw, fill=red!80]
\tikzstyle{asm} = [shape=rectangle, font=\small,
                   minimum width=18em, text width=10em]

\begin{document}

\title{Software Mutational Robustness}

\author{
Eric Schulte\\
\small Computer Science \\
\small U. New Mexico\\
\small Albuquerque, USA\\
\small \t{eschulte@cs.unm.edu}
\and
Zachary P. Fry\\
\small Computer Science\\
\small U. Virginia\\
\small Charlottesville, USA\\
\small \t{zpf5a@cs.virginia.edu}
\and
Ethan Fast\\
\small Computer Science\\
\small Stanford\\
\small Palo Alto, USA\\
\small \t{ethaen@stanford.edu}
\and
Westley Weimer\\
\small Computer Science\\
\small U. Virginia\\
\small Charlottesville, USA \\
\small \t{weimer@cs.virginia.edu}
\and
Stephanie Forrest\\
\begin{tabular}[t]{@{}c@{}}
  \small Computer Science\\
  \small U. New Mexico\\
  \small Albuquerque, USA
\end{tabular}\nobreak\qquad
\begin{tabular}[t]{@{}c@{}}
  \\
  \small Santa Fe Institute\\
  \small Santa Fe, USA
\end{tabular}\\
\small \t{forrest@cs.unm.edu}
}

\date{\today}

\maketitle

\begin{abstract}

  Neutral landscapes and mutational robustness are believed to be
  important enablers of evolvability in biology.  We apply these
  concepts to software, defining \emph{mutational robustness} to be
  the fraction of random mutations to program code that leave a
  program's behavior unchanged.  Test cases are used to measure
  program behavior and mutation operators are taken from earlier work
  on genetic programming.  Although software is often viewed as
  brittle, with small changes leading to catastrophic changes in
  behavior, our results show surprising robustness in the face of
  random software mutations.

  The paper describes empirical studies of the mutational robustness
  of 22 programs, including 14 production software projects, the
  Siemens benchmarks, and four specially constructed programs.  We
  find that over 30\% of random mutations are neutral with respect to
  their test suite.  The results hold across all classes of programs,
  for mutations at both the source code and assembly instruction
  levels, across various programming languages, and bear only a
  limited relation to test suite coverage.  We conclude that
  mutational robustness is an inherent property of software, and that
  neutral variants (i.e., those that pass the test suite) often
  fulfill the program's original purpose or specification.

  Based on these results, we conjecture that neutral mutations can be
  leveraged as a mechanism for generating software diversity.  We
  demonstrate this idea by generating a population of neutral program
  variants and showing that the variants automatically repair latent
  bugs.  Neutral landscapes also provide a partial explanation for
  recent results that use evolutionary computation to automatically
  repair software bugs.
\end{abstract}

\section{Introduction}
\label{sec:intro}

The ability of biological organisms to maintain functionality across a
wide range of environments and to adapt to new environments is
unmatched by engineered systems.  Understanding the intertwined
mechanisms and evolutionary drivers that have led to the robustness
and evolvability of biological systems is an important subfield of
evolutionary biology.  In this paper we focus on neutral landscapes
and mutational robustness, applying these concepts to software.

Today's software arose through fifty years of continued use,
appropriation, and refinement by software developers.  The tools,
design patterns and codes that we have today are those that have
proven useful and were robust to software developer's edits, hacks and
accidents, and those that survived the economic pressures of the
marketplace.  We hypothesize that these evolutionary pressures have
caused software to acquire mutational robustness resembling that of
natural systems.  Mutational robustness in biological systems is
believed to be intimately related to the capacity for unsupervised
evolution and adaptation.  We posit that software mutational
robustness points to the potential for powerful methods of
unsupervised software enhancement and evolution.

Robustness is important in software engineering, especially as it
relates to reliability, availability or dependability.  Here we focus
on genetic robustness, defining software \emph{mutational robustness}
in terms of changes to computer code.  In this context, we define a
\emph{neutral mutation} to be a random change applied to a program
representation (source code, abstract syntax tree, assembly, binary,
etc.) such that the mutated program's behavior remains unchanged on
its regression test suite.\footnote{This definition is not to be
  confused with the ``equivalent mutants'' of mutation testing, see
  Section \ref{sec:mutation-testing}.}  Thus, software fitness is
assessed by the program's performance on a set of test cases.  In
Section \ref{sec:threats-to-validity} we discuss the use of test
suites to assess software fitness.  In Section
\ref{sec:software-mut-rb} we present a formal definition of software
mutational robustness as the fraction of mutations to a program which
do not change its correctness as assessed by a set of regression
tests.

Software mutational robustness measures the fraction of software
mutants that are neutral.  Neutral variants are equivalent to the
original program with respect to the test suite.  They may or may not
be semantically equivalent (compute the same function) to the original
program, they may or may not have the same non-functional properties
(run-time, memory consumption, etc.), and they may or may not satisfy
the specification (required behavior) of the original designers.
Empirically, we find that the program's test suite is an acceptable
proxy for the program specification.  We find many neutral variants
that are both semantically distinct from the original program and
still satisfy the original program's specification or intended
behavior.

This result can be understood more easily when one considers that
there are an infinite number of ways to encode any algorithm in
software.  For example, consider this fragment of a recursive
quick-sort implementation:
\pagebreak
\lstset{language=c}
\begin{lstlisting}
  if (right > left) {
    // code elided ...
    quick(left, r);
    quick(l, right);
  }
\end{lstlisting}
Swapping the order of the last two statements to
\begin{lstlisting}
    quick(l, right);
    quick(left, r);
\end{lstlisting}
changes the run-time behavior of the program without changing the
output, giving an alternate implementation of the specification.  We
find that neutral mutations are prevalent in software and contribute
to evolvability, as discussed in Section \ref{sec:taxonomy}.

Our mutation operators (delete, copy, and swap) are described in
detail in Section \ref{sec:technical-approach}.  They are notable
because they do not create new code de novo.  Delete and copy are both
plausible analogs of genetic operations on DNA, and all three are edit
operations that are routinely performed by programmers.  They are also
related to operators commonly used in the genetic programming
community, although we note that we do not have an explicit terminal
set as typical in genetic programming.  In effect, our terminal set
corresponds to all of the statements contained in the program being
studied.

We are interested in the extent to which mutational robustness enables
software evolvability by which we mean the use of automated methods
for software development and maintenance.  In particular, there is
increasing interest in automatic program repair, and many of the more
promising approaches rely on unsound program transformations (Section
\ref{sec:unsound-transformations}).  These may involve both
source-level edits
(e.g.,~\cite{dallmeier-ase-2009,liblit2011,zeller2010,icse09}) and
modification to program state at run-time
(e.g.,~\cite{rinardClearview}).  Mutational robustness may help
explain why program transformations, such as swapping two
statements~\cite{icse09} or clamping an integer
value~\cite{rinardClearview}, can produce acceptable program behavior.

The primary contributions of the paper include:
\begin{enumerate}

\item The empirical measurement of software mutational robustness in a
  large collection of off-the-shelf software, demonstrating that
  mutational robustness is prevalent.  We find largely uniform
  mutational robustness scores with an average value of $36.8\%$ and a
  minimum across all software instances of $21.2\%$.  We evaluate this
  claim using 22 programs involving over 150,000 lines of code and
  23,151 tests.

\item An application of software mutational robustness to repair
  unknown software defects proactively.  As an illustration, we seeded
  bugs into 11 programs, generated populations 5,000 of neutral
  variants using mutation, and studied the behavior of the variants on
  the seeded bugs using test cases withheld during neutral variant
  generation.  In eight of the programs, the neutral population
  contained at least one variant that ``repaired'' the latent bug,
  passing the withheld test case.

\item An alternate interpretation of the software engineering
  technique known as mutation testing to include neutral mutations and
  software robustness.  We discuss the relation between software
  functionality, software test suites and specifications,
  demonstrating the value of neutral mutations, both for proactively
  repairing unknown bugs and as a likely enabler of automated software
  evolution techniques.

\end{enumerate}

In the remainder of the paper, we review related work in Biology and
Software Engineering in Section \ref{sec:background}.  We then
describe our software representations and mutation operators in
Section \ref{sec:technical-approach}.  The experimental design and
experimental results are given in Section \ref{sec:experimentation}.
We present a practical application of software mutational robustness
in Section \ref{sec:application}.  Finally, we analyze our results and
discuss potential threats to validity and implications in Sections
\ref{sec:discussion} and \ref{sec:conclusion}.

\section{Background}
\label{sec:background}

The previous work most closely related to software mutational
robustness includes work on neutral theories in biology,
investigations of the effect of neutrality in fitness landscapes in
evolutionary computation and the field of mutation testing in software
engineering.  In the following three subsections we highlight some of
the most relevant aspects of these fields.

\subsection{Biology}
\label{sec:background-bio}

Biological evolution is understood in terms of the interplay between
genotype and phenotype.  The genotype is the informational
representation that specifies the organism, and the phenotype is the
physical appearance and behavior of organisms interacting with their
environment.  There is a corresponding type of robustness for each of
these levels of description: \emph{mutational robustness} and
\emph{environmental robustness}
respectively~\cite{kitano2004biological}.  Mutational robustness is
the organism's ability to maintain phenotypic traits in the face of
internal genetic mutations, and environmental robustness is its
ability to maintain functionality across a wide range of
environments~\cite{wagner-rb}.

The two types of robustness are closely related.  Many of the causes
of mutational robustness are also causes of environmental robustness
\cite{lenski2006balancing}.  It is thought that the pervasive
mutational robustness observed in biological systems may have arisen
as a by-product of evolutionary pressure for environmental robustness
\cite{meiklejohn2002single}.  However mutational robustness has been
shown to be beneficial in its own right, especially in its impact on
an organism's evolvability~\cite{ciliberti2007innovation,ofria2008gradual}.

Over time, populations of organisms accumulate mutations in their
genome.  Of the many mutations that occur in a single individual, only
a tiny fraction spread to \emph{fixation} in the population.
Mutations accumulate at a fairly constant rate known as the
\emph{genetic clock}~\cite{zuckerkandl1962molecular}.  Initially, only
those mutations which increased fitness were thought to become fixed
in the population, an idea known as
selectionism~\cite{fisher1930genetical}.  In 1968 Kimura suggested
that because populations have finite size, the majority of accumulated
mutations might be effectively neutral, with no impact upon
fitness~\cite{kimura68}.  Kimura noted that as a consequence of
neutral mutation, ``we must recognize the great importance of random
genetic drift due to finite population number in forming the genetic
structure of biological populations.''

Recent work~\cite{eyre2007distribution} estimates that roughly 50\% of
the fixed mutations provide a selective advantage in Drosophila fruit
flies, which have effective population sizes on the order of $10^6$,
while in hominids, with effective populations sizes of $10^4$, a tiny
percentage of fix mutations were adaptive.  In this study, roughly
16\% of non-equivalent mutations in Drosophila were found to be
effectively neutral compared to roughly 30\% of non-equivalent
mutations in hominids.

The variants of an organism produced through neutral mutations are
called ``neutral neighbors.''  Connected sets of neutral neighbors are
called ``neutral spaces''~\cite{kimura1985neutral} and can occupy
sizeable regions of an organism's fitness landscape
\cite{schuster1994sequences}.  Mutational robustness and the resulting
neutral spaces in fitness landscapes are believed to contribute to a
population's ability to evolve \cite{huynen1996smoothness}.

Mutational drift through neutral spaces gives populations access to
new phenotypes located along the mutational border of the neutral
space.  Neutral spaces thus allow populations to increase diversity
and to accumulate new genetic material through drift.  This
accumulation of genetic material has been shown to be required for
large evolutionary innovations
\cite{ciliberti2007innovation,ofria2008gradual,wagner2008neutralism,meyers2005potential}.

In effect, mutational robustness of an organism is a metric of its
fitness landscape.  A number of metrics of fitness landscapes have
been devised in an attempt to statistically characterize landscapes
\cite{hordijk1996measure} and to directly measure those properties of
a landscape that encourage evolution \cite{smith2002fitness}.  Our
proposed software mutational robustness begins the work of applying
such metrics to real-world software.

\subsection{Evolutionary Computation}
\label{sec:background-ec}

The role of neutrality in evolutionary computation has been explored
in several specific contexts. In simple GP systems whose fitness
landscapes have similarities to those of RNA, neutral mutations have
been shown to enhance exploration in evolutionary searches
\cite{banzhaf2006evolution}.  Neutrality was shown to be beneficial
for evolving digital circuits, both in retrospective analysis of
successful experiments \cite{harvey1997through}, and in directed
experiments using synthetic fitness landscapes designed with variable
amounts of neutrality \cite{vassilev2000advantages}.  Some GP methods
such as Cartesian genetic programming \cite{miller2011cartesian} have
been explicitly designed to leverage neutrality in genetic search.

Using a population genetics model, varying levels of mutational
robustness have been shown to either inhibit or encourage evolution,
depending on population size, mutation rate, and fitness landscape
\cite{draghi2010mutational}.  Some studies (e.g., using a GA to
optimize robot movement \cite{smith2001neutral}) suggest that with
certain complex genotype-phenotype mappings, periods of neutral
evolution do not measurably increase a population's evolvability.
Although none of this prior work studies neutrality in software per
se, it does suggest that success in evolving software (e.g., repairing
bugs) may be related to neutral landscapes in the space of program
representations.

\subsection{Software Engineering}
\label{sec:background-se}

Three subfields of software engineering, namely mutation testing,
n-version programming, and program transformation are most relevant to
this work.

\subsubsection{Mutation Testing}
\label{sec:mutation-testing}

The software engineering community has studied randomly generated
program mutants for over 30 years under the mantle of ``Mutation
Testing''; however, the interpretation and use of program mutants has
been limited to measuring test suite adequacy.  In their landmark
review of mutation testing Jia and Harmon describe the field as
follows \cite{mutation-testing}.

\begin{quotation}
  \noindent
  Mutation Testing is a fault-based testing technique which provides a
  testing criterion called the ``mutation adequacy score.''  The
  mutation adequacy score can be used to measure the effectiveness of
  a test set in terms of its ability to detect [mutants] faults.
\end{quotation}

In mutation testing, non-equivalent mutants are presumed to indicate a
fault (either in the mutant or the original program).  Thus, mutants
that pass a program's test suite indicate test suite failures and
lower the test suite's ``mutation adequacy score.''  Mutation testing
recognizes the existence of ``equivalent mutants'' which are
semantically identical to the original program (cf. the
\emph{Equivalent Mutant Problem}~\cite{budd82}) and viewed as
problematic for the mutation testing paradigm.  The mutation testing
literature, however, does not recognize the possibility of valid
neutral mutants; i.e., mutants that are semantically distinct from the
original program but still satisfy the original program's
specification (and its test suite).

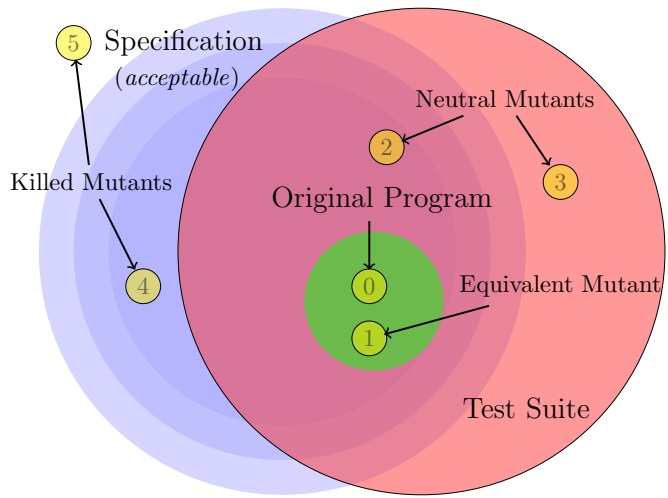
\begin{figure}[htb]
  \centering
  \adjustbox{width=0.75\textwidth}{
    \begin{tikzpicture}[fill opacity=0.5]
      \draw node[font=\LARGE, fill opacity=1] at (1.25,4) {Program Syntactic Space};

      \fill[blue!33]  \speca;
      \fill[blue!33]  \specb;
      \fill[blue!33]  \specc;
      \fill[red]      \test;
      \fill[green]    \equiv;
      \draw       node[label]      at (-1.15,2.75)     {Specification\\ ~{\normalsize (\emph{acceptable})}};
      \draw       node[label] (op) at ( 1.25, 0.75)  {Original Program};
      \draw       node[label]      at ( 4,-2.25) {Test Suite};

      \fill[yellow] \orig;
      \fill[yellow] \muta;
      \fill[yellow] \mutb;
      \fill[yellow] \mutc;
      \fill[yellow] \mutd;
      \fill[yellow] \mute;

      \draw \orig node (orig) {0};
      \draw \muta node (muta) {2};
      \draw \mutb node (mutb) {1};
      \draw \mutc node (mutc) {4};
      \draw \mutd node (mutd) {3};
      \draw \mute node (mute) {5};

      \draw node[class] (neut) at (3.2,2.2) {Neutral Mutants};
      \draw node[class] (equi) at (4,-0.5)  {Equivalent Mutant};
      \draw node[class] (kill) at (-2.75,1) {Killed Mutants};

      \draw[->,thick] (op)   to (orig);
      \draw[->,thick] (neut) to (muta);
      \draw[->,thick] (neut) to (mutd);
      \draw[->,thick] (equi) to (mutb);
      \draw[->,thick] (kill) to (mute);
      \draw[->,thick] (kill) to (mutc);
    \end{tikzpicture}  
  }
  \caption{Syntactic Space of a Program.  \normalfont \small The set
    of programs satisfying the program specification are shaded blue
    (left), the set of programs passing the program's test suite are
    shaded red (right), and the set of equivalent programs are shown
    in green (center).  The relative size of these spaces and of their
    intersection is unknown.  Three classes of mutants are shown and
    labeled.  This work leverages the existence of valid neutral
    mutants such as 2 which lie in the intersection of the
    specification and the test suite.}
  \label{fig:syntactic-space}
\end{figure}

Figure \ref{fig:syntactic-space} shows the syntactic space surrounding
a program. This is similar to a fitness landscape; each point in the
space represents a syntactically distinct program, and each program is
associated with a semantic interpretation although that is not shown
in the figure. Randomly mutating a program's syntactic representation
can have several possible semantic effects, which are shown in the
figure.

Our results are based on the following insight: for every
specification there exists multiple non-equivalent correct
implementations.  This emphasizes a different view of software than is
implicit in the mutation testing technique, namely, that for every
program specification all correct implementations are semantically
equivalent.

To see how this follows from mutation testing, assume that $\exists$
programs \emph{a} and \emph{b} s.t.  \emph{a} is not equivalent to
\emph{b} ($a \not \equiv b$) and both \emph{a} and \emph{b} satisfy
specification S.  Without loss of generality let \emph{a} be the
original program and \emph{b} be a mutant of \emph{a}.  Let T be a
test suite of S.  According to Offut~\cite{offutt1997automatically}
there are two possibilities when T is applied to \emph{b}.  Either
``the mutant is killable, but the test cases is insufficient'' or
``the mutant is functionally equivalent.''  The former case is
impossible because b is assumed to be a correct implementation of S
and thus should not be killed by any test suite of S.  The later case
is impossible because we assume $a \not \equiv b$.  By contraction,
$\forall$ \emph{a} and \emph{b} satisfying the same specification S,
$a \equiv b$ or $\forall$ specification S $\exists! a$ s.t. a
satisfies S.

Taking this perspective, by requiring test suites to kill all
non-equivalent neutral mutants, mutation testing could lead to test
suites that are more restrictive than the specification.  To
illustrate this point, consider the specification and programs in
Figure \ref{fig:spec}.  Programs \emph{a} and \emph{b} both satisfy
the specification \emph{S}, yet they are not equivalent (notably on
many four element arrays).  Any test suite for \emph{S} which kills
one of these implementations will be overly restrictive.

\begin{figure}
  \begin{lstlisting}
/*
 * Spec (S):
 *    Pre: parameter P is an array of three integer elements
 *    Post: returns the smallest of the three input elements
 */

int a(int p[]) {
  if (p[0] <= p[1] && p[0] <= p[2]) return p[0];
  if (p[1] <= p[2] && p[1] <= p[0]) return p[1];
  else return p[2];
}

int b(int p[]) {
  sort(p, "ascending");
  return p[0];
}
  \end{lstlisting}
  \caption{Specification \emph{S} and two correct, non-equivalent
    implementations.}
  \label{fig:spec}
\end{figure}

In the following sections, we study non-equivalent neutral mutants and
investigate their relative frequency in our benchmark programs. Our
study extends work on mutation testing by emphasizing non-equivalent
neutral mutations (cf. neutral mutants 2 and 3 in
Figure~\ref{fig:syntactic-space} which behave identically to the
original program with respect to a given test suite), discussing how
they can be leveraged, and by providing a biological interpretation
for the phenomenon of neutral mutations.

\subsection{N-Version Programming}
\label{sec:n-version}

There has been considerable research on the use of automated diversity
in security, for example, the special issue of \emph{IEEE Computer
  Security} devoted to IT Monocultures~\cite{IEEE-SP-09}.  Common
mechanisms for introducing diversity include Address Space
Randomization \cite{BhatkarEtAl03a,forrest-diversity} and Instruction
Set Randomization \cite{BarrantesEtAl03b,kcetal03a}, among many
others.  In these applications, diversity is introduced to reduce the
risk of widely replicated attacks, by forcing the attacker to redesign
the attack each time it is applied~\cite{strata-diversity}.  Our
proposed use of diversity, outlined in Section \ref{sec:application},
is closer in spirit to \emph{n}-variant systems \cite{nvariant}, where
multiple variants of a program are run in parallel, giving each
variant identical inputs and checking that they all behave similarly
before forwarding the output to the user.

Our proposed application also resembles \emph{n}-version programming
\cite{LittlewoodAndMiller89a} where multiple independent, or
quasi-independent, manually written implementations of critical
programs reduce the risk of implementation errors going undetected.
Our approach differs from \emph{n}-version programming: Instead of
relying on teams of human programmers to generate full implementations
independently, we use lightweight mutation operators to automatically
generate variants that are semantically similar to the original.  This
addresses the cost issue identified in earlier studies
\cite{knightammann89} and potentially addresses the assumption that
independent teams of humans are likely to generate programs that will
fail independently; studies suggest that this assumption does not hold
in practice~\cite{knight-leveson}. Because we generate variants
automatically, there is a better chance of achieving independence
among the variations, either with the mutation operators we describe
here or with others to be developed in the future.

\subsubsection{Unsound Program Transformation}
\label{sec:unsound-transformations}

Traditionally, automated program transformation techniques (e.g.,
compiler optimizations) refrain from altering the semantics (or
behavior) of the original program.  Such program transformations are
``sound'' because they are guaranteed to preserve program semantics.
Recent work has experimented with ``unsound'' program transformations,
which do not necessarily preserve the exact semantics of the original
program.

\emph{Self-healing} systems~\cite{confwoss2002} acknowledge that
proactive developer efforts at providing fault tolerance and
reliability are often insufficient, suggesting instead the idea of
\emph{reactive} methods of runtime protection.  For example, the
Assure system~\cite{assure} adds rescue points to software which catch
otherwise fatal errors, and handle them by re-purposing error handling
code already present in the application.  Similarly, \emph{failure
  oblivious} computing~\cite{failure-oblivious}, handles common memory
errors, such as out-of-bounds reads and writes, by simply ignoring
them or automatically re-mapping invalid memory addresses to arbitrary
valid memory address, allowing the program to continue execution.
Such trade-offs of correctness for robustness are desirable for
applications such as web servers where availability is paramount.

\emph{Loop perforation}~\cite{misailovic2011probabilistically} is a
run-time method that sacrifices exact program accuracy in return for
reduced running time or energy consumption.  Loop perforation
eliminates some of the computation specified in the original program
by dropping some iterations of selected program loops.

Another important class of unsound program transformations are
automated repair techniques.  These techniques share a common
approach: defining a notion of correct and incorrect program behavior
(e.g., from test cases~\cite{icse09,rinardClearview}, implicit
specifications~\cite{liblit2011}, or explicit
specifications~\cite{zeller2010}); generating a set of candidate
repair transformations (e.g., at random~\cite{icse09}, by constraint
solving, or from an established
set~\cite{liblit2011,rinardClearview}), and validating the candidates
produced by the transformations until a suitable repair is found.

Evolutionary methods have been widely used in this work, both to
repair bugs seeded into constructed programs using a subsets of the
Java programming language \cite{arcuri2011evolutionary} and to repair
real programs written in C \cite{icse09,genprog-tse-journal},
including a systematic study of over 100 bugs mined from open-source
software repositories \cite{genprog-icse2012}.  Recent work emphasizes
evolving software patches directly, rather than complete program
representations \cite{ackling11,genprog-icse2012}, evolving repairs at
the assembly and binary code levels \cite{ase10asm,schulte2013}, and
evolving the Java Byte code compiled from extant programs using
specialized mutation operators including a technique of
\emph{compatible crossover}~\cite{orlovandsipper09a,orlovandsipper11}.
Finally, Wilkerson used GP to co-evolve C++ applications in
competition with sets of test cases
\cite{wilkerson2010coevolutionary}.

\section{Technical Approach}
\label{sec:technical-approach}

In this section, we define \emph{software mutational robustness},
describe the program representations used in our experiments, and for
each representation specify the representation-specific mutation
operations.

\subsection{Software Mutational Robustness}
\label{sec:software-mut-rb}

We formalize software mutational robustness with respect to a software
program $P$ (a member of the set of all software programs
$\mathcal{P}$), a set of mutation operators $M$ (where each $m \in M$
is a function mapping $\mathcal{P} \rightarrow \mathcal{P}$), and a
test suite $T : \mathcal{P} \rightarrow \{ \mathsf{true},
\mathsf{false} \}$. A program $P$ is said to pass the test suite iff
$T(P) = \mathsf{true}$.

Given a program $P$, a set of mutation operators $M$, and a test suite
$T$ such that $T(P) = \mathsf{true}$, we define the \emph{software
  mutational robustness}, written $\mathit{MutRB}(P,T,M)$, to be the
fraction of all direct mutants $P'=m(P)$, $\forall m \in M$ which both
compile and pass $T$:

$$
\mathit{MutRB}(P,T,M) =
\frac
{|\{ P' ~|~ m \in M.~ P' = m(P) ~\wedge~ T(P') = \mathsf{true} \}|}
{|\{ P' ~|~ m \in M.~ P' = m(P) \}|}
$$

Based on this definition, software mutational robustness depends on
three parameters, $P$, $T$ and $M$. Perhaps surprisingly, the
empirical results of Section \ref{sec:experimentation} show that
software robustness does not depend strongly on $P$ or $T$ for
human-constructed software systems.  Our mutation operators $M$,
described below, are adapted from genetic programming and are simple
and natural analogs of biological mutation.  We believe that they are
also general and appropriate to software.  For example, earlier work
has shown that the set $M$, together with crossover, is sufficiently
strong to generate successful repairs for a wide variety of defects in
a wide variety of
software~\cite{icse09,genprog-tse-journal,genprog-icse2012}.
Additionally, they reflect common human edit operations.

\subsection{Representation and Operators}
\label{sec:rep-and-ops}

We consider two levels of program representation: abstract syntax
trees (AST) based on high-level program source code, and low-level
assembly code (ASM).  We use the {\sc Cil} toolkit~\cite{necula02cil}
to parse and manipulate ASTs of C source code.  {\sc Cil} simplifies
some C constructs to facilitate manipulation by computational tools
and supports source to source translations such as our mutation
operations.  The ASM representation is the linear sequence of
instructions taken directly from the compiled \texttt{.s} assembly
code file produced by standard compilation (i.e., ``gcc -O2 -S'') on a
64-bit Intel platform, which is split on line breaks~\cite{ase10asm} ,
but with directives and other pseudo-operations protected from
mutation.  Our choice of one tree-based and one linear representation
increases our confidence that the results do not depend on such
representation details. For example, our AST representation is at the
statement level.  That is, each node in the AST corresponds to a legal
statement in C.  This relatively coarse representation level provides
a distinct contrast to the fine-grained ASM representation, where
approximately three assembly statements represent each C statement.

\begin{figure*}
  \centering
  \begin{minipage}[b]{0.28\linewidth}
      \begin{tikzpicture}[scale=0.5]
        \node [gre] at (-2,0) {}
        child { node [gre] {} {
            child { node [yel] {} }
            child { node [gre] {} }}}
        child { node [blu] {}};

        \node [font=\LARGE] at (0,-1.5) {$\rightarrow$};

        \node [gre] at (2,0) {}
        child { node [gre] {} {
            child { node [yel] {} }
            child { node [gre] {} }}}
        child { node [blu] {} {
            child { node [yel] {} }}};
      \end{tikzpicture}
    \subcaption{\small Copy AST}\label{fig:mut-ops-copy-ast}
  \end{minipage}
  \hspace{12pt}
  \begin{minipage}[b]{0.28\linewidth}
      \begin{tikzpicture}[scale=0.5]
        \node [gre] at (-2,0) {}
        child { node [gre] {} {
            child { node [red] {} }
            child { node [gre] {} }}}
        child { node [gre] {}};

        \node [font=\LARGE] at (0,-1.5) {$\rightarrow$};

        \node [gre] at (2,0) {}
        child { node [gre] {} {
            child { node [gre] {} }}}
        child { node [gre] {}};
      \end{tikzpicture}
    \subcaption{\small Delete AST}\label{fig:mut-ops-delete-ast}
  \end{minipage}
  \hspace{12pt}
  \begin{minipage}[b]{0.28\linewidth}
      \begin{tikzpicture}[scale=0.5]
        \node [gre] at (-2,0) {}
        child { node [gre] {} {
            child { node [yel] {} }
            child { node [gre] {} }}}
        child { node [blu] {}};

        \node [font=\LARGE] at (0,-1.5) {$\rightarrow$};

        \node [gre] at (2,0) {}
        child { node [gre] {} {
            child { node [blu] {} }
            child { node [gre] {} }}}
        child { node [yel] {}};
      \end{tikzpicture}
    \subcaption{\small Swap AST}\label{fig:mut-ops-swap-ast}
  \end{minipage}
  \\
  \begin{minipage}[b]{0.32\linewidth}
\adjustbox{width=1.0\linewidth}{
\begin{tikzpicture}[scale=0.4]
\useasboundingbox (-10,5) rectangle (7,-5);
  \node [asm] at (-5,0) {
\begin{verbatim}
movq	8(%rdx), %rdi
xorl	%eax, %eax
\end{verbatim}
\vspace{-16pt}
{\color{orange}
\begin{verbatim}
movq	-80(%rbp), %rdx
\end{verbatim}
}
\vspace{-16pt}
\begin{verbatim}
addl	$1, %r14d
call	atoi
\end{verbatim}
\vspace{-16pt}
{\color{blue}
\begin{verbatim}
movq	-80(%rbp), %rdx
\end{verbatim}
}
\vspace{-16pt}
\begin{verbatim}
movl	%eax, (%r15)
addq	$4, %r15

\end{verbatim}
  };

  \node [font=\LARGE] at (-1.5,0) {$\rightarrow$};

  \node [asm] at (3.5,0) {
\begin{verbatim}
movq	8(%rdx), %rdi
xorl	%eax, %eax
\end{verbatim}
\vspace{-16pt}
{\color{orange}
\begin{verbatim}
movq	-80(%rbp), %rdx
\end{verbatim}
}
\vspace{-16pt}
\begin{verbatim}
addl	$1, %r14d
call	atoi
\end{verbatim}
\vspace{-16pt}
{\color{blue}
\begin{verbatim}
movq	%rdx, -80(%rbp)
\end{verbatim}
}
\vspace{-16pt}
{\color{orange}
\begin{verbatim}
movq	-80(%rbp), %rdx
\end{verbatim}
}
\vspace{-16pt}
\begin{verbatim}
movl	%eax, (%r15)
addq	$4, %r15
\end{verbatim}
};
\end{tikzpicture}
}
    \subcaption{\small Copy ASM}
  \end{minipage}
  \begin{minipage}[b]{0.32\linewidth}
\adjustbox{width=1.0\textwidth}{

\begin{tikzpicture}[scale=0.4]
\useasboundingbox (-10,5) rectangle (7,-5);

  \node [asm] at (-5,0) {
\begin{verbatim}
movq	8(%rdx), %rdi
xorl	%eax, %eax
\end{verbatim}
\vspace{-16pt}
{\color{red}
\begin{verbatim}
movq	%rdx, -80(%rbp)
\end{verbatim}
}
\vspace{-16pt}
\begin{verbatim}
addl	$1, %r14d
call	atoi
movq	-80(%rbp), %rdx
movl	%eax, (%r15)
addq	$4, %r15

\end{verbatim}
  };

  \node [font=\LARGE] at (-1.5,0) {$\rightarrow$};

  \node [asm] at (3.5,0) {
\begin{verbatim}
movq	8(%rdx), %rdi
xorl	%eax, %eax
addl	$1, %r14d
call	atoi
movq	%rdx, -80(%rbp)
movl	%eax, (%r15)
addq	$4, %r15


\end{verbatim}
  };
\end{tikzpicture}
}
    \subcaption{\small Delete ASM}
  \end{minipage}
  \begin{minipage}[b]{0.32\linewidth}
    \adjustbox{width=1.0\textwidth}{
\begin{tikzpicture}[scale=0.4]
\useasboundingbox (-10,5) rectangle (7,-5);

  \node [asm] at (-5,0) {
\begin{verbatim}
movq	8(%rdx), %rdi
xorl	%eax, %eax
\end{verbatim}
\vspace{-16pt}
{\color{orange}
\begin{verbatim}
movq	%rdx, -80(%rbp)
\end{verbatim}
}
\vspace{-16pt}
\begin{verbatim}
addl	$1, %r14d
call	atoi
\end{verbatim}
\vspace{-16pt}
{\color{blue}
\begin{verbatim}
movq	-80(%rbp), %rdx
\end{verbatim}
}
\vspace{-16pt}
\begin{verbatim}
movl	%eax, (%r15)
addq	$4, %r15

\end{verbatim}
  };

  \node [font=\LARGE] at (-1.5,0) {$\rightarrow$};

  \node [asm] at (3.5,0) {
\begin{verbatim}
movq	8(%rdx), %rdi
xorl	%eax, %eax
\end{verbatim}
\vspace{-16pt}
{\color{blue}
\begin{verbatim}
movq	-80(%rbp), %rdx
\end{verbatim}
}
\vspace{-16pt}
\begin{verbatim}
addl	$1, %r14d
call	atoi
\end{verbatim}
\vspace{-16pt}
{\color{orange}
\begin{verbatim}
movq	%rdx, -80(%rbp)
\end{verbatim}
}
\vspace{-16pt}
\begin{verbatim}
movl	%eax, (%r15)
addq	$4, %r15

\end{verbatim}
  };
\end{tikzpicture}
}
    \subcaption{\small Swap ASM}
  \end{minipage}
  \caption{Mutation operators: \small{Copy, Delete, Swap.}\label{fig:mut-ops}}
\end{figure*}
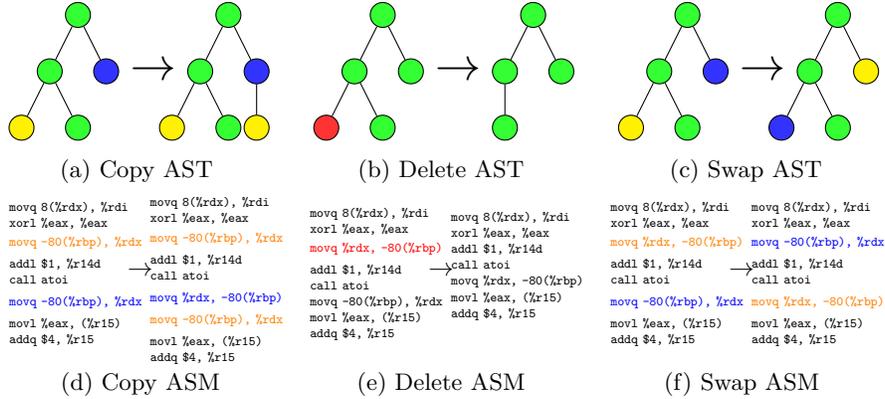

Given a source code or assembly language program, we consider three
simple language-independent mutation operators: \emph{copy},
\emph{delete} and \emph{swap}.  Copy duplicates an AST statement-level
subtree or assembly instruction and inserts it at a random position in
the AST or immediately after a randomly chosen instruction.  Delete
removes a randomly chosen statement-level AST subtree or assembly
instruction.  Swap exchanges two randomly chosen statement-level AST
subtrees or assembly instructions.  Figure \ref{fig:mut-ops}
illustrates these operators.  Because AST mutations manipulate
subtrees, a large amount of code might be inserted or deleted by a
single mutation, depending on how high in the tree the mutation is
applied.  In the experiments, mutations modify only AST statements or
ASM instructions that are actually visited by the test suite.  This
restriction is similar to mutating only those parts of genome that are
known to be involved in the phenotype being assayed.  Mutations to
untested statements would likely be neutral under our metric, unfairly
biasing the results towards overly high estimates of mutational
robustness.

\section{Experimental Results}
\label{sec:experimentation}

We report results for five experiments on the mutational robustness of
programs in both representations (AST and ASM).  Throughout this
section we remain conscious of the threat that our results may measure
the poor quality of program test suites instead of the intrinsic
mutational robustness of software.  Much of our experimental design
addresses this particular concern.

We investigate: (1) the level of mutational robustness of several
computer programs (we specifically select benchmark programs with the
highest quality test suites available); (2) the extent to which
mutational robustness depends on or can be explained by test suite
quality; (3) a taxonomy of neutral mutations (indicating which
mutations would no longer remain neutral under perfect testing), (4)
the effect of cumulative neutral mutations mutations in a simple
program, and (5) generality of mutational robustness across multiple
programming languages and paradigms.

\subsection{Benchmark Programs}
\label{sec:benchmark}

\begin{table*}
\begin{center}
\begin{tabular}{lrrrrrrl}
 Program & \multicolumn{2}{c}{Lines of Code} & \multicolumn{2}{c}{Test Suite} & \multicolumn{2}{c}{Mut. Robustness} \\
                                                 & ASM        & C          & \# Tests  & \% Stmt. & AST                & ASM                \\
\toprule
Sorting Algorithms                               &            &            &           &          &                    &                    \\
\midrule
 bubble-sort                                     & 184        & 34         & 10        & 100      & 27.3               & 25.7               \\
 insertion-sort                                  & 170        & 29         & 10        & 100      & 29.4               & 26.0               \\
 merge-sort                                      & 233        & 38         & 10        & 100      & 29.8               & 21.2               \\
 quick-sort                                      & 219        & 38         & 10        & 100      & 28.9               & 25.5               \\
\midrule
Siemens~\cite{hutchins1994experiments}$\dagger$ &            &            &           &          &                    &                    \\
\midrule
 printtokens                                     & 2419       & 536        & 4130      & 81.7     & 21.2               & 25.8               \\
 schedule                                        & 922        & 412        & 2650      & 94.4     & 34.4               & 29.1               \\
 space                                           & 18098      & 9126       & 13494     & 91.1     & 37.7               & 32.1               \\
 tcas                                            & 544        & 173        & 1608      & 96.2     & 33.5               & 25.9               \\
\midrule
Systems Programs                                 &            &            &           &          &                    &                    \\
\midrule
 bzip2 1.0.2                                     & 18756      & 7000       & 6         & 35.9     & 33.0               & 26.1               \\
 --- \emph{(alt. test suite)}                    &            &            & 22        & 71.0     & 46.4               & 23.6               \\
 ccrypt 1.2                                      & 15261      & 4249       & 6         & 29.5     & 33.0               & 69.7               \\
 --- \emph{(alt. test suite)}                    &            &            & 16        & 40.4     & 34.6               & 69.7               \\
 grep                                            & 28776      & 10929      & 119       & 24.9     & 50.0               & 36.7               \\
 imagemagick 6.5.2                               & 6128       & 147        & 145       & 0.8      & 33.3               & 66.3               \\
 jansson 1.3                                     & 6830       & 2975       & 30        & 28.8     & 33.3               & 28.0               \\
 leukocyte                                       & 40226      & 7970       & 5         & 45.4     & 33.3               & 39.9               \\
 lighttpd 1.4.15                                 & 34165      & 3829       & 11        & 40.1     & 61.5               & 56.9               \\
 nullhttpd 0.5.0                                 & 5951       & 5575       & 6         & 64.5     & 41.5               & 37.8               \\
 oggenc 1.0.1                                    & 299959     & 59094      & 10        & 38.4     & 33.4               & 22.1               \\
 --- \emph{(alt. test suite)}                    &            &            & 40        & 58.8     & 40.5               & 72.3               \\
 potion 40b5f03                                  & 80406      & 15033      & 204       & 48.4     & 33.3               & 48.9               \\
 redis 1.3.4                                     & 44802      & 17203      & 234       & 9.2      & 33.3               & 34.0               \\
 sed                                             & 17026      & 8059       & 360       & 42.0     & 33.0               & 25.6               \\
 tiff 3.8.2                                      & 22458      & 1732       & 10        & 15.4     & 33.3               & 90.4               \\
 vyquon 335426d                                  & 20567      & 4390       & 5         & 50.6     & 33.3               & 69.0               \\
\midrule
 \bf total or average                            & \bf 664100 & \bf 158571 & \bf 23151 & \bf 40.9 & \bf 33.9 $\pm10$ & \bf 39.6 $\pm22$ \\
\bottomrule
\end{tabular}


\caption{Mutational robustness of benchmark programs.  \small ``Lines
  of Code'' columns report the size of the program in terms of lines
  of C source code and lines of compiled assembly code. The ``Test
  Suite'' columns show the size of the test suite both in terms of
  number of test cases and the percentage of all AST level statements
  in the program that are exercised by the test suite.  The
  ``Mut. Robustness'' columns report the percentage of all first-order
  mutations that were neutral.  The $\pm$ values in the bottom row
  indicate one standard deviation.  For each program, at both the AST
  and ASM level we generated at least 200 unique variants using each
  of the three mutation operations (\emph{copy}, \emph{delete} and
  \emph{swap}).  Mutation operations were applied at locations chosen
  randomly from all those visited by the test cases. For three
  programs (\texttt{bzip}, \texttt{ccrypt} and \texttt{oggenc}) we
  also evaluated on three independent alternate test suites.\newline
  $\dagger$ Although the Siemens benchmark suite claims complete
  branch and statement coverage, we find less than $100\%$ statement
  coverage.  This is due to our use of finer-grained Cil statements in
  calculating coverage.  See Section \ref{sec:rep-and-ops} for
  discussion of the Cil program
  representation. \label{tab:benchmarks-rb}}
\end{center}
\end{table*}

We selected 22 programs for our experiments (Table
\ref{tab:benchmarks-rb}).  Fourteen are off-the-shelf programs
selected to measure mutational robustness in real-world software.
Four are taken from the Siemens Software-artifact Infrastructure
Repository, created by Siemens Research \cite{hutchins1994experiments}
and later modified by Rothermel and Harold
\cite{rothermel1998empirical} until each ``executable statement, edge,
and definition-use pair in the base program or its control flow graph
was exercised by at least 30 tests.''  The \texttt{space} test suite,
which was generated by Vokolos~\cite{vokolos1998empirical} and later
enhanced by Graves~\cite{graves2001empirical}, which covers every edge
in the control flow graph with at least 30 tests.  These programs are
included for comparability to previous research and to study the
robustness of programs with extremely high-quality test suites.  We
include four simple sorting algorithms taken from
\url{http://rosettacode.org} to demonstrate the robustness of programs
with full statement, branch-level and assembly instruction test
coverage.

Each program has an associated test suite. The tests either came with
the program as part of its established regression test (e.g., Siemens,
\texttt{potion}, \texttt{redis}, \texttt{jansson}) or were constructed
manually (e.g., sorting algorithms, webservers).  A number of our
benchmarks implement invertible transformations (e.g., compression,
encryption, serialization, image manipulation), which form an implicit
formal specification and permit simple testing~\cite{wah00}.  The
three invertible programs (\texttt{bzip}, \texttt{ccrypt} and
\texttt{oggenc}) are thus each evaluated on two independently
constructed test suites. For \texttt{lighttpd} and
\texttt{imagemagick}, we restricted mutations to
\texttt{mod\_fastcgi.c} and \texttt{convert.c} respectively,
demonstrating application to modules as well as to whole systems.

\subsection{Software Mutational Robustness}
\label{sec:software-mutational-robustness}

We first demonstrate that a variety of software programs exhibit
significant mutational robustness under the mutation operators
described in Section \ref{sec:technical-approach}. In this experiment,
we measure the percentage of random mutations to code visited by at
least one test case which leave the program's behavior unchanged on
all of its test cases.  In every case, the initial program passes all
of its test cases.  For each mutation operator we generate program
variations making at least 200 copies of the original program and then
applying a single random mutation to each copy.  We refer to these as
\emph{first-order}, mutations.  We then run the mutated program on its
test suite and count it as neutral if it passes all of its tests.

Table \ref{tab:benchmarks-rb} shows the results of this experiment on
the benchmark programs.  We wish to rule out trivial mutations, such
as the insertion of dead code (e.g., statements that appear after a
\texttt{return}) or the transposition of independent lines, that are
visible in the source code but would produce equivalent assembly code.
Since program equivalence is undecidable~~\cite{budd82}, we
approximate this by compiling the AST using ``gcc~-O2~-S,'' which
includes dead code elimination, SSA form, and instruction scheduling.
Multiple source code variants that produce the same optimized assembly
code (modulo label names and other directives) are counted only once
in Table \ref{tab:benchmarks-rb}.  Similarly, any two ASM-level
mutations which produce the same executable are counted only once.

Although the results vary by program (e.g., \texttt{grep} is more
robust than \texttt{printtokens}), the results show a remarkably high
level of mutational robustness: Across all programs, operators, and
representations (source or assembly) 36.8\% of variants continue to
pass all test cases with no systematic difference between the AST and
ASM representation.  In the next two subsections we ask to what extent
these results arise from inadequate test suites
(\ref{sec:test-suite-quality}) or from semantically equivalent
mutations (\ref{sec:taxonomy}).

\subsection{Does Mutational Robustness Depend on Test Suite Quality?}
\label{sec:test-suite-quality}

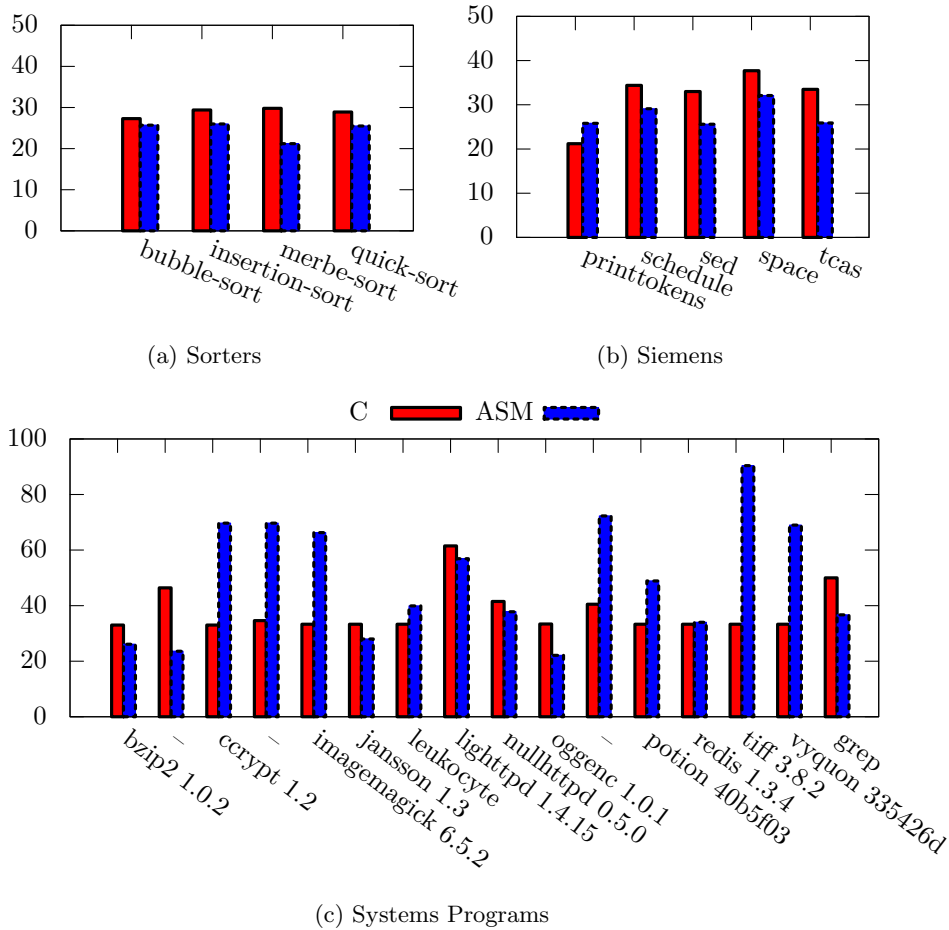
\begin{figure}
  \centering
  \begin{minipage}[b]{0.49\linewidth}
\begin{tikzpicture}[gnuplot]
\path (0.000,0.000) rectangle (6.500,4.750);
\gpcolor{color=gp lt color border}
\gpsetlinetype{gp lt border}
\gpsetlinewidth{1.00}
\draw[gp path] (1.012,1.272)--(1.192,1.272);
\draw[gp path] (5.697,1.272)--(5.517,1.272);
\node[gp node right] at (0.828,1.272) { 0};
\draw[gp path] (1.012,1.819)--(1.192,1.819);
\draw[gp path] (5.697,1.819)--(5.517,1.819);
\node[gp node right] at (0.828,1.819) { 10};
\draw[gp path] (1.012,2.366)--(1.192,2.366);
\draw[gp path] (5.697,2.366)--(5.517,2.366);
\node[gp node right] at (0.828,2.366) { 20};
\draw[gp path] (1.012,2.913)--(1.192,2.913);
\draw[gp path] (5.697,2.913)--(5.517,2.913);
\node[gp node right] at (0.828,2.913) { 30};
\draw[gp path] (1.012,3.460)--(1.192,3.460);
\draw[gp path] (5.697,3.460)--(5.517,3.460);
\node[gp node right] at (0.828,3.460) { 40};
\draw[gp path] (1.012,4.007)--(1.192,4.007);
\draw[gp path] (5.697,4.007)--(5.517,4.007);
\node[gp node right] at (0.828,4.007) { 50};
\draw[gp path] (1.949,1.272)--(1.949,1.452);
\draw[gp path] (1.949,4.007)--(1.949,3.827);
\node[gp node left,rotate=-22] at (1.949,1.088) {bubble-sort};
\draw[gp path] (2.886,1.272)--(2.886,1.452);
\draw[gp path] (2.886,4.007)--(2.886,3.827);
\node[gp node left,rotate=-22] at (2.886,1.088) {insertion-sort};
\draw[gp path] (3.823,1.272)--(3.823,1.452);
\draw[gp path] (3.823,4.007)--(3.823,3.827);
\node[gp node left,rotate=-22] at (3.823,1.088) {merbe-sort};
\draw[gp path] (4.760,1.272)--(4.760,1.452);
\draw[gp path] (4.760,4.007)--(4.760,3.827);
\node[gp node left,rotate=-22] at (4.760,1.088) {quick-sort};
\draw[gp path] (1.012,4.007)--(1.012,1.272)--(5.697,1.272)--(5.697,4.007)--cycle;
\gpfill{color=gp lt color 0} (1.832,1.272)--(2.067,1.272)--(2.067,2.766)--(1.832,2.766)--cycle;
\gpsetlinetype{gp lt plot 0}
\gpsetlinewidth{3.00}
\draw[gp path] (1.832,1.272)--(1.832,2.765)--(2.066,2.765)--(2.066,1.272)--cycle;
\gpfill{color=gp lt color 0} (2.769,1.272)--(3.004,1.272)--(3.004,2.881)--(2.769,2.881)--cycle;
\draw[gp path] (2.769,1.272)--(2.769,2.880)--(3.003,2.880)--(3.003,1.272)--cycle;
\gpfill{color=gp lt color 0} (3.706,1.272)--(3.941,1.272)--(3.941,2.903)--(3.706,2.903)--cycle;
\draw[gp path] (3.706,1.272)--(3.706,2.902)--(3.940,2.902)--(3.940,1.272)--cycle;
\gpfill{color=gp lt color 0} (4.643,1.272)--(4.878,1.272)--(4.878,2.854)--(4.643,2.854)--cycle;
\draw[gp path] (4.643,1.272)--(4.643,2.853)--(4.877,2.853)--(4.877,1.272)--cycle;
\gpfill{color=gp lt color 2} (2.066,1.272)--(2.301,1.272)--(2.301,2.679)--(2.066,2.679)--cycle;
\gpsetlinetype{gp lt plot 2}
\draw[gp path] (2.066,1.272)--(2.066,2.678)--(2.300,2.678)--(2.300,1.272)--cycle;
\gpfill{color=gp lt color 2} (3.003,1.272)--(3.238,1.272)--(3.238,2.695)--(3.003,2.695)--cycle;
\draw[gp path] (3.003,1.272)--(3.003,2.694)--(3.237,2.694)--(3.237,1.272)--cycle;
\gpfill{color=gp lt color 2} (3.940,1.272)--(4.175,1.272)--(4.175,2.433)--(3.940,2.433)--cycle;
\draw[gp path] (3.940,1.272)--(3.940,2.432)--(4.174,2.432)--(4.174,1.272)--cycle;
\gpfill{color=gp lt color 2} (4.877,1.272)--(5.112,1.272)--(5.112,2.668)--(4.877,2.668)--cycle;
\draw[gp path] (4.877,1.272)--(4.877,2.667)--(5.111,2.667)--(5.111,1.272)--cycle;
\gpsetlinetype{gp lt border}
\gpsetlinewidth{1.00}
\draw[gp path] (1.012,4.007)--(1.012,1.272)--(5.697,1.272)--(5.697,4.007)--cycle;
\gpdefrectangularnode{gp plot 1}{\pgfpoint{1.012cm}{1.272cm}}{\pgfpoint{5.697cm}{4.007cm}}
\end{tikzpicture}
    \subcaption{\small Sorters\label{rb-class-sort}}
  \end{minipage}
  \begin{minipage}[b]{0.49\linewidth}
\begin{tikzpicture}[gnuplot]
\path (0.000,0.000) rectangle (6.500,4.750);
\gpcolor{color=gp lt color border}
\gpsetlinetype{gp lt border}
\gpsetlinewidth{1.00}
\draw[gp path] (1.012,1.066)--(1.192,1.066);
\draw[gp path] (5.697,1.066)--(5.517,1.066);
\node[gp node right] at (0.828,1.066) { 0};
\draw[gp path] (1.012,1.654)--(1.192,1.654);
\draw[gp path] (5.697,1.654)--(5.517,1.654);
\node[gp node right] at (0.828,1.654) { 10};
\draw[gp path] (1.012,2.242)--(1.192,2.242);
\draw[gp path] (5.697,2.242)--(5.517,2.242);
\node[gp node right] at (0.828,2.242) { 20};
\draw[gp path] (1.012,2.831)--(1.192,2.831);
\draw[gp path] (5.697,2.831)--(5.517,2.831);
\node[gp node right] at (0.828,2.831) { 30};
\draw[gp path] (1.012,3.419)--(1.192,3.419);
\draw[gp path] (5.697,3.419)--(5.517,3.419);
\node[gp node right] at (0.828,3.419) { 40};
\draw[gp path] (1.012,4.007)--(1.192,4.007);
\draw[gp path] (5.697,4.007)--(5.517,4.007);
\node[gp node right] at (0.828,4.007) { 50};
\draw[gp path] (1.793,1.066)--(1.793,1.246);
\draw[gp path] (1.793,4.007)--(1.793,3.827);
\node[gp node left,rotate=-22] at (1.793,0.882) {printtokens};
\draw[gp path] (2.574,1.066)--(2.574,1.246);
\draw[gp path] (2.574,4.007)--(2.574,3.827);
\node[gp node left,rotate=-22] at (2.574,0.882) {schedule};
\draw[gp path] (3.355,1.066)--(3.355,1.246);
\draw[gp path] (3.355,4.007)--(3.355,3.827);
\node[gp node left,rotate=-22] at (3.355,0.882) {sed};
\draw[gp path] (4.135,1.066)--(4.135,1.246);
\draw[gp path] (4.135,4.007)--(4.135,3.827);
\node[gp node left,rotate=-22] at (4.135,0.882) {space};
\draw[gp path] (4.916,1.066)--(4.916,1.246);
\draw[gp path] (4.916,4.007)--(4.916,3.827);
\node[gp node left,rotate=-22] at (4.916,0.882) {tcas};
\draw[gp path] (1.012,4.007)--(1.012,1.066)--(5.697,1.066)--(5.697,4.007)--cycle;
\gpfill{color=gp lt color 0} (1.695,1.066)--(1.891,1.066)--(1.891,2.314)--(1.695,2.314)--cycle;
\gpsetlinetype{gp lt plot 0}
\gpsetlinewidth{3.00}
\draw[gp path] (1.695,1.066)--(1.695,2.313)--(1.890,2.313)--(1.890,1.066)--cycle;
\gpfill{color=gp lt color 0} (2.476,1.066)--(2.672,1.066)--(2.672,3.090)--(2.476,3.090)--cycle;
\draw[gp path] (2.476,1.066)--(2.476,3.089)--(2.671,3.089)--(2.671,1.066)--cycle;
\gpfill{color=gp lt color 0} (3.257,1.066)--(3.453,1.066)--(3.453,3.008)--(3.257,3.008)--cycle;
\draw[gp path] (3.257,1.066)--(3.257,3.007)--(3.452,3.007)--(3.452,1.066)--cycle;
\gpfill{color=gp lt color 0} (4.038,1.066)--(4.234,1.066)--(4.234,3.285)--(4.038,3.285)--cycle;
\draw[gp path] (4.038,1.066)--(4.038,3.284)--(4.233,3.284)--(4.233,1.066)--cycle;
\gpfill{color=gp lt color 0} (4.819,1.066)--(5.015,1.066)--(5.015,3.037)--(4.819,3.037)--cycle;
\draw[gp path] (4.819,1.066)--(4.819,3.036)--(5.014,3.036)--(5.014,1.066)--cycle;
\gpfill{color=gp lt color 2} (1.890,1.066)--(2.087,1.066)--(2.087,2.585)--(1.890,2.585)--cycle;
\gpsetlinetype{gp lt plot 2}
\draw[gp path] (1.890,1.066)--(1.890,2.584)--(2.086,2.584)--(2.086,1.066)--cycle;
\gpfill{color=gp lt color 2} (2.671,1.066)--(2.867,1.066)--(2.867,2.779)--(2.671,2.779)--cycle;
\draw[gp path] (2.671,1.066)--(2.671,2.778)--(2.866,2.778)--(2.866,1.066)--cycle;
\gpfill{color=gp lt color 2} (3.452,1.066)--(3.648,1.066)--(3.648,2.573)--(3.452,2.573)--cycle;
\draw[gp path] (3.452,1.066)--(3.452,2.572)--(3.647,2.572)--(3.647,1.066)--cycle;
\gpfill{color=gp lt color 2} (4.233,1.066)--(4.429,1.066)--(4.429,2.955)--(4.233,2.955)--cycle;
\draw[gp path] (4.233,1.066)--(4.233,2.954)--(4.428,2.954)--(4.428,1.066)--cycle;
\gpfill{color=gp lt color 2} (5.014,1.066)--(5.210,1.066)--(5.210,2.590)--(5.014,2.590)--cycle;
\draw[gp path] (5.014,1.066)--(5.014,2.589)--(5.209,2.589)--(5.209,1.066)--cycle;
\gpsetlinetype{gp lt border}
\gpsetlinewidth{1.00}
\draw[gp path] (1.012,4.007)--(1.012,1.066)--(5.697,1.066)--(5.697,4.007)--cycle;
\gpdefrectangularnode{gp plot 1}{\pgfpoint{1.012cm}{1.066cm}}{\pgfpoint{5.697cm}{4.007cm}}
\end{tikzpicture}
    \subcaption{\small Siemens\label{rb-class-siemens}}
  \end{minipage}
  \\
  \begin{minipage}[b]{1.0\linewidth}
\begin{tikzpicture}[gnuplot]
\path (0.000,0.000) rectangle (12.500,6.750);
\gpcolor{color=gp lt color border}
\gpsetlinetype{gp lt border}
\gpsetlinewidth{1.00}
\draw[gp path] (1.196,2.190)--(1.376,2.190);
\draw[gp path] (11.947,2.190)--(11.767,2.190);
\node[gp node right] at (1.012,2.190) { 0};
\draw[gp path] (1.196,2.929)--(1.376,2.929);
\draw[gp path] (11.947,2.929)--(11.767,2.929);
\node[gp node right] at (1.012,2.929) { 20};
\draw[gp path] (1.196,3.668)--(1.376,3.668);
\draw[gp path] (11.947,3.668)--(11.767,3.668);
\node[gp node right] at (1.012,3.668) { 40};
\draw[gp path] (1.196,4.408)--(1.376,4.408);
\draw[gp path] (11.947,4.408)--(11.767,4.408);
\node[gp node right] at (1.012,4.408) { 60};
\draw[gp path] (1.196,5.147)--(1.376,5.147);
\draw[gp path] (11.947,5.147)--(11.767,5.147);
\node[gp node right] at (1.012,5.147) { 80};
\draw[gp path] (1.196,5.886)--(1.376,5.886);
\draw[gp path] (11.947,5.886)--(11.767,5.886);
\node[gp node right] at (1.012,5.886) { 100};
\draw[gp path] (1.828,2.190)--(1.828,2.370);
\draw[gp path] (1.828,5.886)--(1.828,5.706);
\node[gp node left,rotate=-37] at (1.828,2.006) {bzip2 1.0.2};
\draw[gp path] (2.461,2.190)--(2.461,2.370);
\draw[gp path] (2.461,5.886)--(2.461,5.706);
\node[gp node left,rotate=-37] at (2.461,2.006) {--};
\draw[gp path] (3.093,2.190)--(3.093,2.370);
\draw[gp path] (3.093,5.886)--(3.093,5.706);
\node[gp node left,rotate=-37] at (3.093,2.006) {ccrypt 1.2};
\draw[gp path] (3.726,2.190)--(3.726,2.370);
\draw[gp path] (3.726,5.886)--(3.726,5.706);
\node[gp node left,rotate=-37] at (3.726,2.006) {--};
\draw[gp path] (4.358,2.190)--(4.358,2.370);
\draw[gp path] (4.358,5.886)--(4.358,5.706);
\node[gp node left,rotate=-37] at (4.358,2.006) {imagemagick 6.5.2};
\draw[gp path] (4.990,2.190)--(4.990,2.370);
\draw[gp path] (4.990,5.886)--(4.990,5.706);
\node[gp node left,rotate=-37] at (4.990,2.006) {jansson 1.3};
\draw[gp path] (5.623,2.190)--(5.623,2.370);
\draw[gp path] (5.623,5.886)--(5.623,5.706);
\node[gp node left,rotate=-37] at (5.623,2.006) {leukocyte};
\draw[gp path] (6.255,2.190)--(6.255,2.370);
\draw[gp path] (6.255,5.886)--(6.255,5.706);
\node[gp node left,rotate=-37] at (6.255,2.006) {lighttpd 1.4.15};
\draw[gp path] (6.888,2.190)--(6.888,2.370);
\draw[gp path] (6.888,5.886)--(6.888,5.706);
\node[gp node left,rotate=-37] at (6.888,2.006) {nullhttpd 0.5.0};
\draw[gp path] (7.520,2.190)--(7.520,2.370);
\draw[gp path] (7.520,5.886)--(7.520,5.706);
\node[gp node left,rotate=-37] at (7.520,2.006) {oggenc 1.0.1};
\draw[gp path] (8.153,2.190)--(8.153,2.370);
\draw[gp path] (8.153,5.886)--(8.153,5.706);
\node[gp node left,rotate=-37] at (8.153,2.006) {--};
\draw[gp path] (8.785,2.190)--(8.785,2.370);
\draw[gp path] (8.785,5.886)--(8.785,5.706);
\node[gp node left,rotate=-37] at (8.785,2.006) {potion 40b5f03};
\draw[gp path] (9.417,2.190)--(9.417,2.370);
\draw[gp path] (9.417,5.886)--(9.417,5.706);
\node[gp node left,rotate=-37] at (9.417,2.006) {redis 1.3.4};
\draw[gp path] (10.050,2.190)--(10.050,2.370);
\draw[gp path] (10.050,5.886)--(10.050,5.706);
\node[gp node left,rotate=-37] at (10.050,2.006) {tiff 3.8.2};
\draw[gp path] (10.682,2.190)--(10.682,2.370);
\draw[gp path] (10.682,5.886)--(10.682,5.706);
\node[gp node left,rotate=-37] at (10.682,2.006) {vyquon 335426d};
\draw[gp path] (11.315,2.190)--(11.315,2.370);
\draw[gp path] (11.315,5.886)--(11.315,5.706);
\node[gp node left,rotate=-37] at (11.315,2.006) {grep};
\draw[gp path] (1.196,5.886)--(1.196,2.190)--(11.947,2.190)--(11.947,5.886)--cycle;
\node[gp node right] at (5.287,6.228) {C};
\gpfill{color=gp lt color 0} (5.471,6.151)--(6.387,6.151)--(6.387,6.305)--(5.471,6.305)--cycle;
\gpsetlinetype{gp lt plot 0}
\gpsetlinewidth{3.00}
\draw[gp path] (5.471,6.151)--(6.387,6.151)--(6.387,6.305)--(5.471,6.305)--cycle;
\gpfill{color=gp lt color 0} (1.749,2.190)--(1.908,2.190)--(1.908,3.411)--(1.749,3.411)--cycle;
\draw[gp path] (1.749,2.190)--(1.749,3.410)--(1.907,3.410)--(1.907,2.190)--cycle;
\gpfill{color=gp lt color 0} (2.382,2.190)--(2.541,2.190)--(2.541,3.906)--(2.382,3.906)--cycle;
\draw[gp path] (2.382,2.190)--(2.382,3.905)--(2.540,3.905)--(2.540,2.190)--cycle;
\gpfill{color=gp lt color 0} (3.014,2.190)--(3.173,2.190)--(3.173,3.411)--(3.014,3.411)--cycle;
\draw[gp path] (3.014,2.190)--(3.014,3.410)--(3.172,3.410)--(3.172,2.190)--cycle;
\gpfill{color=gp lt color 0} (3.647,2.190)--(3.806,2.190)--(3.806,3.470)--(3.647,3.470)--cycle;
\draw[gp path] (3.647,2.190)--(3.647,3.469)--(3.805,3.469)--(3.805,2.190)--cycle;
\gpfill{color=gp lt color 0} (4.279,2.190)--(4.438,2.190)--(4.438,3.422)--(4.279,3.422)--cycle;
\draw[gp path] (4.279,2.190)--(4.279,3.421)--(4.437,3.421)--(4.437,2.190)--cycle;
\gpfill{color=gp lt color 0} (4.911,2.190)--(5.071,2.190)--(5.071,3.422)--(4.911,3.422)--cycle;
\draw[gp path] (4.911,2.190)--(4.911,3.421)--(5.070,3.421)--(5.070,2.190)--cycle;
\gpfill{color=gp lt color 0} (5.544,2.190)--(5.703,2.190)--(5.703,3.422)--(5.544,3.422)--cycle;
\draw[gp path] (5.544,2.190)--(5.544,3.421)--(5.702,3.421)--(5.702,2.190)--cycle;
\gpfill{color=gp lt color 0} (6.176,2.190)--(6.335,2.190)--(6.335,4.464)--(6.176,4.464)--cycle;
\draw[gp path] (6.176,2.190)--(6.176,4.463)--(6.334,4.463)--(6.334,2.190)--cycle;
\gpfill{color=gp lt color 0} (6.809,2.190)--(6.968,2.190)--(6.968,3.725)--(6.809,3.725)--cycle;
\draw[gp path] (6.809,2.190)--(6.809,3.724)--(6.967,3.724)--(6.967,2.190)--cycle;
\gpfill{color=gp lt color 0} (7.441,2.190)--(7.600,2.190)--(7.600,3.425)--(7.441,3.425)--cycle;
\draw[gp path] (7.441,2.190)--(7.441,3.424)--(7.599,3.424)--(7.599,2.190)--cycle;
\gpfill{color=gp lt color 0} (8.073,2.190)--(8.233,2.190)--(8.233,3.688)--(8.073,3.688)--cycle;
\draw[gp path] (8.073,2.190)--(8.073,3.687)--(8.232,3.687)--(8.232,2.190)--cycle;
\gpfill{color=gp lt color 0} (8.706,2.190)--(8.865,2.190)--(8.865,3.422)--(8.706,3.422)--cycle;
\draw[gp path] (8.706,2.190)--(8.706,3.421)--(8.864,3.421)--(8.864,2.190)--cycle;
\gpfill{color=gp lt color 0} (9.338,2.190)--(9.497,2.190)--(9.497,3.422)--(9.338,3.422)--cycle;
\draw[gp path] (9.338,2.190)--(9.338,3.421)--(9.496,3.421)--(9.496,2.190)--cycle;
\gpfill{color=gp lt color 0} (9.971,2.190)--(10.130,2.190)--(10.130,3.422)--(9.971,3.422)--cycle;
\draw[gp path] (9.971,2.190)--(9.971,3.421)--(10.129,3.421)--(10.129,2.190)--cycle;
\gpfill{color=gp lt color 0} (10.603,2.190)--(10.762,2.190)--(10.762,3.422)--(10.603,3.422)--cycle;
\draw[gp path] (10.603,2.190)--(10.603,3.421)--(10.761,3.421)--(10.761,2.190)--cycle;
\gpfill{color=gp lt color 0} (11.236,2.190)--(11.395,2.190)--(11.395,4.039)--(11.236,4.039)--cycle;
\draw[gp path] (11.236,2.190)--(11.236,4.038)--(11.394,4.038)--(11.394,2.190)--cycle;
\node[gp node right] at (7.5,6.228) {ASM};
\gpfill{color=gp lt color 2} (7.5,6.151)--(8.223,6.151)--(8.223,6.305)--(7.5,6.305)--cycle;
\gpsetlinetype{gp lt plot 2}
\draw[gp path] (7.5,6.151)--(8.223,6.151)--(8.223,6.305)--(7.5,6.305)--cycle;
\gpfill{color=gp lt color 2} (1.907,2.190)--(2.067,2.190)--(2.067,3.156)--(1.907,3.156)--cycle;
\draw[gp path] (1.907,2.190)--(1.907,3.155)--(2.066,3.155)--(2.066,2.190)--cycle;
\gpfill{color=gp lt color 2} (2.540,2.190)--(2.699,2.190)--(2.699,3.063)--(2.540,3.063)--cycle;
\draw[gp path] (2.540,2.190)--(2.540,3.062)--(2.698,3.062)--(2.698,2.190)--cycle;
\gpfill{color=gp lt color 2} (3.172,2.190)--(3.331,2.190)--(3.331,4.767)--(3.172,4.767)--cycle;
\draw[gp path] (3.172,2.190)--(3.172,4.766)--(3.330,4.766)--(3.330,2.190)--cycle;
\gpfill{color=gp lt color 2} (3.805,2.190)--(3.964,2.190)--(3.964,4.767)--(3.805,4.767)--cycle;
\draw[gp path] (3.805,2.190)--(3.805,4.766)--(3.963,4.766)--(3.963,2.190)--cycle;
\gpfill{color=gp lt color 2} (4.437,2.190)--(4.596,2.190)--(4.596,4.641)--(4.437,4.641)--cycle;
\draw[gp path] (4.437,2.190)--(4.437,4.640)--(4.595,4.640)--(4.595,2.190)--cycle;
\gpfill{color=gp lt color 2} (5.070,2.190)--(5.229,2.190)--(5.229,3.226)--(5.070,3.226)--cycle;
\draw[gp path] (5.070,2.190)--(5.070,3.225)--(5.228,3.225)--(5.228,2.190)--cycle;
\gpfill{color=gp lt color 2} (5.702,2.190)--(5.861,2.190)--(5.861,3.666)--(5.702,3.666)--cycle;
\draw[gp path] (5.702,2.190)--(5.702,3.665)--(5.860,3.665)--(5.860,2.190)--cycle;
\gpfill{color=gp lt color 2} (6.334,2.190)--(6.493,2.190)--(6.493,4.294)--(6.334,4.294)--cycle;
\draw[gp path] (6.334,2.190)--(6.334,4.293)--(6.492,4.293)--(6.492,2.190)--cycle;
\gpfill{color=gp lt color 2} (6.967,2.190)--(7.126,2.190)--(7.126,3.588)--(6.967,3.588)--cycle;
\draw[gp path] (6.967,2.190)--(6.967,3.587)--(7.125,3.587)--(7.125,2.190)--cycle;
\gpfill{color=gp lt color 2} (7.599,2.190)--(7.758,2.190)--(7.758,3.008)--(7.599,3.008)--cycle;
\draw[gp path] (7.599,2.190)--(7.599,3.007)--(7.757,3.007)--(7.757,2.190)--cycle;
\gpfill{color=gp lt color 2} (8.232,2.190)--(8.391,2.190)--(8.391,4.863)--(8.232,4.863)--cycle;
\draw[gp path] (8.232,2.190)--(8.232,4.862)--(8.390,4.862)--(8.390,2.190)--cycle;
\gpfill{color=gp lt color 2} (8.864,2.190)--(9.023,2.190)--(9.023,3.998)--(8.864,3.998)--cycle;
\draw[gp path] (8.864,2.190)--(8.864,3.997)--(9.022,3.997)--(9.022,2.190)--cycle;
\gpfill{color=gp lt color 2} (9.496,2.190)--(9.656,2.190)--(9.656,3.448)--(9.496,3.448)--cycle;
\draw[gp path] (9.496,2.190)--(9.496,3.447)--(9.655,3.447)--(9.655,2.190)--cycle;
\gpfill{color=gp lt color 2} (10.129,2.190)--(10.288,2.190)--(10.288,5.532)--(10.129,5.532)--cycle;
\draw[gp path] (10.129,2.190)--(10.129,5.531)--(10.287,5.531)--(10.287,2.190)--cycle;
\gpfill{color=gp lt color 2} (10.761,2.190)--(10.920,2.190)--(10.920,4.741)--(10.761,4.741)--cycle;
\draw[gp path] (10.761,2.190)--(10.761,4.740)--(10.919,4.740)--(10.919,2.190)--cycle;
\gpfill{color=gp lt color 2} (11.394,2.190)--(11.553,2.190)--(11.553,3.547)--(11.394,3.547)--cycle;
\draw[gp path] (11.394,2.190)--(11.394,3.546)--(11.552,3.546)--(11.552,2.190)--cycle;
\gpsetlinetype{gp lt border}
\gpsetlinewidth{1.00}
\draw[gp path] (1.196,5.886)--(1.196,2.190)--(11.947,2.190)--(11.947,5.886)--cycle;
\gpdefrectangularnode{gp plot 1}{\pgfpoint{1.196cm}{2.190cm}}{\pgfpoint{11.947cm}{5.886cm}}
\end{tikzpicture}
    \subcaption{\small Systems Programs\label{rb-class-open}}
  \end{minipage}
  \caption{Mutational robustness shown by quality of test suite.
    \small The simple sorting programs in \ref{rb-class-sort} have
    complete code and ASM instruction coverage with few carefully
    constructed test cases.  The Siemens programs in
    \ref{rb-class-siemens} have complete branch and def-use pair
    coverage and thousands of test cases for relatively small
    programs.  The Systems programs in \ref{rb-class-open} have test
    suites that vary greatly in quality.\label{rb-classes}}
\end{figure}

The results in Table \ref{tab:benchmarks-rb} are striking, but they
could potentially be entirely the result of inadequate test suites.
In Figure \ref{rb-classes} we plot the data from Table
\ref{tab:benchmarks-rb}, showing the mutational robustness of software
grouped by the quality of the software test suite.  We consider both
the quantitative metrics of code coverage of these test suites, as
well as qualitative differences between the program test suites by
panel.

Qualitatively the program classes in the three panels of Figure
\ref{rb-classes} have very different test suites.

\begin{description}
\item[\textbf{Sorting}] program suites all share a single test suite.
  This test suite leverages the simplicity of sorting programs to
  provide complete code coverage (at both the statement and assembly
  instruction levels) with only 10 test cases.
\item[\textbf{Siemens}] program suites taken from the testing
  community where they have been developed by multiple parties across
  multiple publications
  \cite{hutchins1994experiments,rothermel1998empirical} until each
  ``executable statement, edge, and definition-use pair in the base
  program or its control flow graph was exercised by at least 30
  tests.''
\item[\textbf{Systems}] program suites are taken directly from real
  world software development projects.  These test suites reflect the
  great range of test suites used in practice.
\end{description}

Quantitatively, the sorting programs in Panel \ref{rb-class-sort} have
100\% code coverage and the Siemens benchmark programs in Panel
\ref{rb-class-siemens} have close to 100\% code coverage.  By contrast
the Systems programs in Panel \ref{rb-class-open} have $37.63\pm19.34$
coverage.  Despite this wide range of coverage, the average mutational
robustness varies relatively little between Panels at 26.7\% (Panel
\ref{rb-class-sort}), 29.8\% (Panel \ref{rb-class-siemens}) and 43.7\%
(Panel \ref{rb-class-open}) respectively, and the minimum mutational
robustness measurements for each Panel are even closer at 21.2\%,
21.2\% and 22.1\% respectively.

The lack of correlation between code coverage and mutational
robustness is not surprising, both because we explicitly limit the
mutation operators to code that is visited by test suites, and because
statement coverage is known to be an insufficient metric of test suite
quality~\cite{howden76}.

In fact, the lack of correlation between mutational robustness and
test suite quality is not true in either limit.  At one extreme, even
high-quality test suites (such as the Siemens benchmarks, which were
explicitly designed to test all execution paths) and test suites with
full statement, branch and assembly instruction coverage have over
20\% mutational robustness.  At the other extreme, a minimal test
suite that we designed for bubble sort, which does not check program
output but requires only successful compilation and execution without
crash, has only 84.8\% mutational robustness.

These results show that, in practice, for both real programs and
comprehensively tested ones, mutational robustness is not fully
explained by test suite quality or inadequacy.

\subsection{Taxonomy of Neutral Variants}
\label{sec:taxonomy}

To provide insight into our results, we chose bubble sort as an
example of an easy-to-understand and easy-to-test program.  We then
studied the effect of 35 first-order neutral mutations on bubble sort
at the AST level, developing a taxonomy of the neutral mutations.
After manual review, all 35 neutral variations were confirmed to be
valid implementations of the sorting specification.  We next
categorized them with respect to their operational differences from
the original.  The results are shown in Table \ref{tab:taxonomy}.

\begin{table}
  \centering
  \begin{tabular}{r|l|r}
    \toprule
    \# & Functional Category                          & Frequency/35 \\
    \toprule
    1  & Different whitespace in output               & 12           \\
    2  & Inconsequential change of internal variables & 10           \\
    3  & Extra or redundant computation               & 6            \\
    4  & Equivalent or redundant conditional guard    & 3            \\
    5  & Switched to non-explicit return              & 2            \\
    6  & Changed code is unreachable                  & 1            \\
    7  & Removed optimization                         & 1            \\
    \bottomrule
  \end{tabular}
  \caption{Taxonomy of 35 neutral first order AST variations of
    bubble sort.  \small Categorized by manual
    review. ``Different whitespace in output'' describes variants
    whose output differs from the original program only in whitespace
    characters which are not detected by the test suite.
    ``Inconsequential change of internal variables'' describes variants whose
    mutations change the behavior of the program while executing but
    do not change the tested output of the program, e.g., mutations which
    change the values of variables in memory which do not later affect
    program output or variants which change the ordering of
    non-interacting instructions.}
  \label{tab:taxonomy}
\end{table}

These categories have different effects on program execution.  Only
categories 1 and 5 affect the externally observable behavior of the
program by changing output and return values in ways not specified by
the program specification.  Categories 2, 3, 4 and 7 may affect the
running time of the program.  Category 2 includes the removal of
unnecessary variable assignments, re-ordering non-interacting
instructions and changing state which is later overwritten or never
again read.  Many changes, such as types 2, 3 and 4, produce programs
that will likely be more robust to further manipulation by inserting
redundant (occasionally diverse) control flow guards (i.e.,
conditionals that control if statements) and variable assignments.
Across most of these categories we find alternative implementations
that are \emph{not} semantically equivalent to the original but which
conform to the program specification.

Of these classes of neutral programs, only classes 4, 5 and 6 could
possibly have no impact on runtime behavior and could possibly be
equivalent.  Under the mutation testing paradigm, tests would be
constructed to ``kill'' the remaining 29 of 35 neutral mutants.  Given
the sorting specification used (namely to print whitespace separated
integer inputs in sorted order to \texttt{STDOUT} separated by
whitespace), \emph{none} of these classes of neutral mutations could
be viewed as faulty implementations. Consequently, any such extra tests
constructed to tell them apart (e.g., for mutation testing) would
\emph{over-specify} the program specification.  Rather than improving the
test suite quality, such over-constrained tests would potentially judge
future correct implementations as faulty.

\subsection{Cumulative Robustness}
\label{sec:cumulative}

\begin{figure}
  \begin{minipage}[b]{1.0\linewidth}
    \adjustbox{width=1.0\textwidth}{
\begin{tikzpicture}[gnuplot]
\path (0.000,0.000) rectangle (12.700,7.620);
\gpcolor{color=gp lt color border}
\gpsetlinetype{gp lt border}
\gpsetlinewidth{1.00}
\draw[gp path] (1.504,0.985)--(1.684,0.985);
\node[gp node right] at (1.320,0.985) { 170};
\draw[gp path] (1.504,2.029)--(1.684,2.029);
\node[gp node right] at (1.320,2.029) { 180};
\draw[gp path] (1.504,3.074)--(1.684,3.074);
\node[gp node right] at (1.320,3.074) { 190};
\draw[gp path] (1.504,4.118)--(1.684,4.118);
\node[gp node right] at (1.320,4.118) { 200};
\draw[gp path] (1.504,5.162)--(1.684,5.162);
\node[gp node right] at (1.320,5.162) { 210};
\draw[gp path] (1.504,6.207)--(1.684,6.207);
\node[gp node right] at (1.320,6.207) { 220};
\draw[gp path] (1.504,7.251)--(1.684,7.251);
\node[gp node right] at (1.320,7.251) { 230};
\draw[gp path] (1.504,0.985)--(1.504,1.165);
\draw[gp path] (1.504,7.251)--(1.504,7.071);
\node[gp node center] at (1.504,0.677) { 0};
\draw[gp path] (3.387,0.985)--(3.387,1.165);
\draw[gp path] (3.387,7.251)--(3.387,7.071);
\node[gp node center] at (3.387,0.677) { 50};
\draw[gp path] (5.270,0.985)--(5.270,1.165);
\draw[gp path] (5.270,7.251)--(5.270,7.071);
\node[gp node center] at (5.270,0.677) { 100};
\draw[gp path] (7.153,0.985)--(7.153,1.165);
\draw[gp path] (7.153,7.251)--(7.153,7.071);
\node[gp node center] at (7.153,0.677) { 150};
\draw[gp path] (9.036,0.985)--(9.036,1.165);
\draw[gp path] (9.036,7.251)--(9.036,7.071);
\node[gp node center] at (9.036,0.677) { 200};
\draw[gp path] (10.919,0.985)--(10.919,1.165);
\draw[gp path] (10.919,7.251)--(10.919,7.071);
\node[gp node center] at (10.919,0.677) { 250};
\draw[gp path] (10.919,0.985)--(10.739,0.985);
\node[gp node left] at (11.103,0.985) { 12};
\draw[gp path] (10.919,1.681)--(10.739,1.681);
\node[gp node left] at (11.103,1.681) { 13};
\draw[gp path] (10.919,2.377)--(10.739,2.377);
\node[gp node left] at (11.103,2.377) { 14};
\draw[gp path] (10.919,3.074)--(10.739,3.074);
\node[gp node left] at (11.103,3.074) { 15};
\draw[gp path] (10.919,3.770)--(10.739,3.770);
\node[gp node left] at (11.103,3.770) { 16};
\draw[gp path] (10.919,4.466)--(10.739,4.466);
\node[gp node left] at (11.103,4.466) { 17};
\draw[gp path] (10.919,5.162)--(10.739,5.162);
\node[gp node left] at (11.103,5.162) { 18};
\draw[gp path] (10.919,5.859)--(10.739,5.859);
\node[gp node left] at (11.103,5.859) { 19};
\draw[gp path] (10.919,6.555)--(10.739,6.555);
\node[gp node left] at (11.103,6.555) { 20};
\draw[gp path] (10.919,7.251)--(10.739,7.251);
\node[gp node left] at (11.103,7.251) { 21};
\draw[gp path] (1.504,7.251)--(1.504,0.985)--(10.919,0.985)--(10.919,7.251)--cycle;
\node[gp node center,rotate=-270] at (0.246,4.118) {Avg. LOC};
\node[gp node center,rotate=-270] at (11.992,4.118) {\% Neutral Variants};
\node[gp node center] at (6.211,0.215) {Number of Applied Mutations};
\node[gp node right] at (9.451,1.627) {Avg. LOC};
\gpcolor{color=gp lt color 7}
\gpsetpointsize{4.00}
\gppoint{gp mark 8}{(1.542,1.548)}
\gppoint{gp mark 8}{(1.579,1.580)}
\gppoint{gp mark 8}{(1.617,1.616)}
\gppoint{gp mark 8}{(1.655,1.625)}
\gppoint{gp mark 8}{(1.692,1.661)}
\gppoint{gp mark 8}{(1.730,1.706)}
\gppoint{gp mark 8}{(1.768,1.730)}
\gppoint{gp mark 8}{(1.805,1.725)}
\gppoint{gp mark 8}{(1.843,1.743)}
\gppoint{gp mark 8}{(1.881,1.778)}
\gppoint{gp mark 8}{(1.918,1.801)}
\gppoint{gp mark 8}{(1.956,1.801)}
\gppoint{gp mark 8}{(1.994,1.836)}
\gppoint{gp mark 8}{(2.031,1.851)}
\gppoint{gp mark 8}{(2.069,1.866)}
\gppoint{gp mark 8}{(2.107,1.897)}
\gppoint{gp mark 8}{(2.144,1.933)}
\gppoint{gp mark 8}{(2.182,1.920)}
\gppoint{gp mark 8}{(2.220,1.911)}
\gppoint{gp mark 8}{(2.257,1.938)}
\gppoint{gp mark 8}{(2.295,1.981)}
\gppoint{gp mark 8}{(2.333,2.014)}
\gppoint{gp mark 8}{(2.370,2.053)}
\gppoint{gp mark 8}{(2.408,2.087)}
\gppoint{gp mark 8}{(2.446,2.120)}
\gppoint{gp mark 8}{(2.483,2.093)}
\gppoint{gp mark 8}{(2.521,2.157)}
\gppoint{gp mark 8}{(2.558,2.160)}
\gppoint{gp mark 8}{(2.596,2.147)}
\gppoint{gp mark 8}{(2.634,2.154)}
\gppoint{gp mark 8}{(2.671,2.111)}
\gppoint{gp mark 8}{(2.709,2.115)}
\gppoint{gp mark 8}{(2.747,2.088)}
\gppoint{gp mark 8}{(2.784,2.156)}
\gppoint{gp mark 8}{(2.822,2.168)}
\gppoint{gp mark 8}{(2.860,2.205)}
\gppoint{gp mark 8}{(2.897,2.214)}
\gppoint{gp mark 8}{(2.935,2.282)}
\gppoint{gp mark 8}{(2.973,2.282)}
\gppoint{gp mark 8}{(3.010,2.339)}
\gppoint{gp mark 8}{(3.048,2.423)}
\gppoint{gp mark 8}{(3.086,2.461)}
\gppoint{gp mark 8}{(3.123,2.489)}
\gppoint{gp mark 8}{(3.161,2.556)}
\gppoint{gp mark 8}{(3.199,2.523)}
\gppoint{gp mark 8}{(3.236,2.543)}
\gppoint{gp mark 8}{(3.274,2.567)}
\gppoint{gp mark 8}{(3.312,2.546)}
\gppoint{gp mark 8}{(3.349,2.510)}
\gppoint{gp mark 8}{(3.387,2.505)}
\gppoint{gp mark 8}{(3.425,2.545)}
\gppoint{gp mark 8}{(3.462,2.542)}
\gppoint{gp mark 8}{(3.500,2.551)}
\gppoint{gp mark 8}{(3.538,2.552)}
\gppoint{gp mark 8}{(3.575,2.600)}
\gppoint{gp mark 8}{(3.613,2.671)}
\gppoint{gp mark 8}{(3.651,2.660)}
\gppoint{gp mark 8}{(3.688,2.660)}
\gppoint{gp mark 8}{(3.726,2.554)}
\gppoint{gp mark 8}{(3.764,2.597)}
\gppoint{gp mark 8}{(3.801,2.618)}
\gppoint{gp mark 8}{(3.839,2.548)}
\gppoint{gp mark 8}{(3.877,2.567)}
\gppoint{gp mark 8}{(3.914,2.568)}
\gppoint{gp mark 8}{(3.952,2.555)}
\gppoint{gp mark 8}{(3.990,2.518)}
\gppoint{gp mark 8}{(4.027,2.503)}
\gppoint{gp mark 8}{(4.065,2.470)}
\gppoint{gp mark 8}{(4.103,2.557)}
\gppoint{gp mark 8}{(4.140,2.540)}
\gppoint{gp mark 8}{(4.178,2.486)}
\gppoint{gp mark 8}{(4.216,2.506)}
\gppoint{gp mark 8}{(4.253,2.545)}
\gppoint{gp mark 8}{(4.291,2.543)}
\gppoint{gp mark 8}{(4.329,2.555)}
\gppoint{gp mark 8}{(4.366,2.551)}
\gppoint{gp mark 8}{(4.404,2.607)}
\gppoint{gp mark 8}{(4.441,2.694)}
\gppoint{gp mark 8}{(4.479,2.838)}
\gppoint{gp mark 8}{(4.517,2.819)}
\gppoint{gp mark 8}{(4.554,2.867)}
\gppoint{gp mark 8}{(4.592,2.996)}
\gppoint{gp mark 8}{(4.630,3.017)}
\gppoint{gp mark 8}{(4.667,3.108)}
\gppoint{gp mark 8}{(4.705,3.081)}
\gppoint{gp mark 8}{(4.743,3.094)}
\gppoint{gp mark 8}{(4.780,3.132)}
\gppoint{gp mark 8}{(4.818,3.081)}
\gppoint{gp mark 8}{(4.856,3.089)}
\gppoint{gp mark 8}{(4.893,3.118)}
\gppoint{gp mark 8}{(4.931,3.135)}
\gppoint{gp mark 8}{(4.969,3.115)}
\gppoint{gp mark 8}{(5.006,3.113)}
\gppoint{gp mark 8}{(5.044,3.173)}
\gppoint{gp mark 8}{(5.082,3.059)}
\gppoint{gp mark 8}{(5.119,3.192)}
\gppoint{gp mark 8}{(5.157,3.108)}
\gppoint{gp mark 8}{(5.195,3.160)}
\gppoint{gp mark 8}{(5.232,3.159)}
\gppoint{gp mark 8}{(5.270,3.254)}
\gppoint{gp mark 8}{(5.308,3.321)}
\gppoint{gp mark 8}{(5.345,3.360)}
\gppoint{gp mark 8}{(5.383,3.389)}
\gppoint{gp mark 8}{(5.421,3.385)}
\gppoint{gp mark 8}{(5.458,3.389)}
\gppoint{gp mark 8}{(5.496,3.390)}
\gppoint{gp mark 8}{(5.534,3.394)}
\gppoint{gp mark 8}{(5.571,3.412)}
\gppoint{gp mark 8}{(5.609,3.353)}
\gppoint{gp mark 8}{(5.647,3.375)}
\gppoint{gp mark 8}{(5.684,3.338)}
\gppoint{gp mark 8}{(5.722,3.360)}
\gppoint{gp mark 8}{(5.760,3.370)}
\gppoint{gp mark 8}{(5.797,3.399)}
\gppoint{gp mark 8}{(5.835,3.399)}
\gppoint{gp mark 8}{(5.873,3.445)}
\gppoint{gp mark 8}{(5.910,3.503)}
\gppoint{gp mark 8}{(5.948,3.532)}
\gppoint{gp mark 8}{(5.986,3.533)}
\gppoint{gp mark 8}{(6.023,3.617)}
\gppoint{gp mark 8}{(6.061,3.553)}
\gppoint{gp mark 8}{(6.099,3.566)}
\gppoint{gp mark 8}{(6.136,3.617)}
\gppoint{gp mark 8}{(6.174,3.662)}
\gppoint{gp mark 8}{(6.212,3.717)}
\gppoint{gp mark 8}{(6.249,3.744)}
\gppoint{gp mark 8}{(6.287,3.773)}
\gppoint{gp mark 8}{(6.324,3.842)}
\gppoint{gp mark 8}{(6.362,3.881)}
\gppoint{gp mark 8}{(6.400,3.919)}
\gppoint{gp mark 8}{(6.437,3.935)}
\gppoint{gp mark 8}{(6.475,3.893)}
\gppoint{gp mark 8}{(6.513,3.918)}
\gppoint{gp mark 8}{(6.550,3.935)}
\gppoint{gp mark 8}{(6.588,4.016)}
\gppoint{gp mark 8}{(6.626,4.059)}
\gppoint{gp mark 8}{(6.663,4.118)}
\gppoint{gp mark 8}{(6.701,4.122)}
\gppoint{gp mark 8}{(6.739,4.132)}
\gppoint{gp mark 8}{(6.776,4.165)}
\gppoint{gp mark 8}{(6.814,4.196)}
\gppoint{gp mark 8}{(6.852,4.269)}
\gppoint{gp mark 8}{(6.889,4.265)}
\gppoint{gp mark 8}{(6.927,4.276)}
\gppoint{gp mark 8}{(6.965,4.338)}
\gppoint{gp mark 8}{(7.002,4.380)}
\gppoint{gp mark 8}{(7.040,4.471)}
\gppoint{gp mark 8}{(7.078,4.510)}
\gppoint{gp mark 8}{(7.115,4.525)}
\gppoint{gp mark 8}{(7.153,4.515)}
\gppoint{gp mark 8}{(7.191,4.538)}
\gppoint{gp mark 8}{(7.228,4.563)}
\gppoint{gp mark 8}{(7.266,4.567)}
\gppoint{gp mark 8}{(7.304,4.609)}
\gppoint{gp mark 8}{(7.341,4.601)}
\gppoint{gp mark 8}{(7.379,4.665)}
\gppoint{gp mark 8}{(7.417,4.619)}
\gppoint{gp mark 8}{(7.454,4.609)}
\gppoint{gp mark 8}{(7.492,4.681)}
\gppoint{gp mark 8}{(7.530,4.658)}
\gppoint{gp mark 8}{(7.567,4.688)}
\gppoint{gp mark 8}{(7.605,4.794)}
\gppoint{gp mark 8}{(7.643,4.833)}
\gppoint{gp mark 8}{(7.680,4.891)}
\gppoint{gp mark 8}{(7.718,4.905)}
\gppoint{gp mark 8}{(7.756,4.925)}
\gppoint{gp mark 8}{(7.793,4.939)}
\gppoint{gp mark 8}{(7.831,4.978)}
\gppoint{gp mark 8}{(7.869,5.009)}
\gppoint{gp mark 8}{(7.906,4.945)}
\gppoint{gp mark 8}{(7.944,4.992)}
\gppoint{gp mark 8}{(7.982,5.101)}
\gppoint{gp mark 8}{(8.019,5.084)}
\gppoint{gp mark 8}{(8.057,5.035)}
\gppoint{gp mark 8}{(8.094,5.113)}
\gppoint{gp mark 8}{(8.132,4.983)}
\gppoint{gp mark 8}{(8.170,4.932)}
\gppoint{gp mark 8}{(8.207,4.951)}
\gppoint{gp mark 8}{(8.245,4.876)}
\gppoint{gp mark 8}{(8.283,4.819)}
\gppoint{gp mark 8}{(8.320,4.859)}
\gppoint{gp mark 8}{(8.358,5.005)}
\gppoint{gp mark 8}{(8.396,4.997)}
\gppoint{gp mark 8}{(8.433,5.000)}
\gppoint{gp mark 8}{(8.471,4.981)}
\gppoint{gp mark 8}{(8.509,4.990)}
\gppoint{gp mark 8}{(8.546,5.037)}
\gppoint{gp mark 8}{(8.584,4.945)}
\gppoint{gp mark 8}{(8.622,5.033)}
\gppoint{gp mark 8}{(8.659,5.090)}
\gppoint{gp mark 8}{(8.697,5.135)}
\gppoint{gp mark 8}{(8.735,5.236)}
\gppoint{gp mark 8}{(8.772,5.219)}
\gppoint{gp mark 8}{(8.810,5.257)}
\gppoint{gp mark 8}{(8.848,5.363)}
\gppoint{gp mark 8}{(8.885,5.356)}
\gppoint{gp mark 8}{(8.923,5.404)}
\gppoint{gp mark 8}{(8.961,5.364)}
\gppoint{gp mark 8}{(8.998,5.383)}
\gppoint{gp mark 8}{(9.036,5.412)}
\gppoint{gp mark 8}{(9.074,5.446)}
\gppoint{gp mark 8}{(9.111,5.424)}
\gppoint{gp mark 8}{(9.149,5.345)}
\gppoint{gp mark 8}{(9.187,5.380)}
\gppoint{gp mark 8}{(9.224,5.335)}
\gppoint{gp mark 8}{(9.262,5.406)}
\gppoint{gp mark 8}{(9.300,5.417)}
\gppoint{gp mark 8}{(9.337,5.350)}
\gppoint{gp mark 8}{(9.375,5.439)}
\gppoint{gp mark 8}{(9.413,5.428)}
\gppoint{gp mark 8}{(9.450,5.480)}
\gppoint{gp mark 8}{(9.488,5.518)}
\gppoint{gp mark 8}{(9.526,5.564)}
\gppoint{gp mark 8}{(9.563,5.639)}
\gppoint{gp mark 8}{(9.601,5.651)}
\gppoint{gp mark 8}{(9.639,5.680)}
\gppoint{gp mark 8}{(9.676,5.784)}
\gppoint{gp mark 8}{(9.714,5.827)}
\gppoint{gp mark 8}{(9.752,5.883)}
\gppoint{gp mark 8}{(9.789,5.952)}
\gppoint{gp mark 8}{(9.827,5.996)}
\gppoint{gp mark 8}{(9.865,6.070)}
\gppoint{gp mark 8}{(9.902,6.075)}
\gppoint{gp mark 8}{(9.940,6.053)}
\gppoint{gp mark 8}{(9.978,6.042)}
\gppoint{gp mark 8}{(10.015,6.037)}
\gppoint{gp mark 8}{(10.053,5.990)}
\gppoint{gp mark 8}{(10.090,5.997)}
\gppoint{gp mark 8}{(10.128,5.992)}
\gppoint{gp mark 8}{(10.166,6.010)}
\gppoint{gp mark 8}{(10.203,6.078)}
\gppoint{gp mark 8}{(10.241,6.112)}
\gppoint{gp mark 8}{(10.279,6.113)}
\gppoint{gp mark 8}{(10.316,6.218)}
\gppoint{gp mark 8}{(10.354,6.296)}
\gppoint{gp mark 8}{(10.392,6.358)}
\gppoint{gp mark 8}{(10.429,6.369)}
\gppoint{gp mark 8}{(10.467,6.512)}
\gppoint{gp mark 8}{(10.505,6.608)}
\gppoint{gp mark 8}{(10.542,6.629)}
\gppoint{gp mark 8}{(10.580,6.646)}
\gppoint{gp mark 8}{(10.618,6.673)}
\gppoint{gp mark 8}{(10.655,6.657)}
\gppoint{gp mark 8}{(10.693,6.747)}
\gppoint{gp mark 8}{(10.731,6.810)}
\gppoint{gp mark 8}{(10.093,1.627)}
\gpcolor{color=gp lt color border}
\node[gp node right] at (9.451,1.319) {\% Neutral Variants};
\gpcolor{rgb color={0.000,0.000,1.000}}
\gppoint{gp mark 1}{(1.542,2.634)}
\gppoint{gp mark 1}{(1.579,2.591)}
\gppoint{gp mark 1}{(1.617,3.163)}
\gppoint{gp mark 1}{(1.655,3.227)}
\gppoint{gp mark 1}{(1.692,1.672)}
\gppoint{gp mark 1}{(1.730,2.899)}
\gppoint{gp mark 1}{(1.768,4.215)}
\gppoint{gp mark 1}{(1.805,2.914)}
\gppoint{gp mark 1}{(1.843,3.629)}
\gppoint{gp mark 1}{(1.881,2.839)}
\gppoint{gp mark 1}{(1.918,3.022)}
\gppoint{gp mark 1}{(1.956,3.276)}
\gppoint{gp mark 1}{(1.994,3.375)}
\gppoint{gp mark 1}{(2.031,2.706)}
\gppoint{gp mark 1}{(2.069,2.314)}
\gppoint{gp mark 1}{(2.107,2.869)}
\gppoint{gp mark 1}{(2.144,2.914)}
\gppoint{gp mark 1}{(2.182,3.770)}
\gppoint{gp mark 1}{(2.220,3.988)}
\gppoint{gp mark 1}{(2.257,3.441)}
\gppoint{gp mark 1}{(2.295,3.577)}
\gppoint{gp mark 1}{(2.333,2.591)}
\gppoint{gp mark 1}{(2.370,4.007)}
\gppoint{gp mark 1}{(2.408,3.292)}
\gppoint{gp mark 1}{(2.446,1.840)}
\gppoint{gp mark 1}{(2.483,3.594)}
\gppoint{gp mark 1}{(2.521,2.520)}
\gppoint{gp mark 1}{(2.558,3.664)}
\gppoint{gp mark 1}{(2.596,2.534)}
\gppoint{gp mark 1}{(2.634,4.195)}
\gppoint{gp mark 1}{(2.671,2.464)}
\gppoint{gp mark 1}{(2.709,3.227)}
\gppoint{gp mark 1}{(2.747,3.509)}
\gppoint{gp mark 1}{(2.784,3.391)}
\gppoint{gp mark 1}{(2.822,2.884)}
\gppoint{gp mark 1}{(2.860,3.053)}
\gppoint{gp mark 1}{(2.897,2.750)}
\gppoint{gp mark 1}{(2.935,3.492)}
\gppoint{gp mark 1}{(2.973,4.760)}
\gppoint{gp mark 1}{(3.010,2.991)}
\gppoint{gp mark 1}{(3.048,3.325)}
\gppoint{gp mark 1}{(3.086,3.408)}
\gppoint{gp mark 1}{(3.123,3.577)}
\gppoint{gp mark 1}{(3.161,2.013)}
\gppoint{gp mark 1}{(3.199,2.750)}
\gppoint{gp mark 1}{(3.236,5.289)}
\gppoint{gp mark 1}{(3.274,4.253)}
\gppoint{gp mark 1}{(3.312,2.794)}
\gppoint{gp mark 1}{(3.349,2.677)}
\gppoint{gp mark 1}{(3.387,3.084)}
\gppoint{gp mark 1}{(3.425,3.647)}
\gppoint{gp mark 1}{(3.462,3.458)}
\gppoint{gp mark 1}{(3.500,3.037)}
\gppoint{gp mark 1}{(3.538,3.037)}
\gppoint{gp mark 1}{(3.575,3.116)}
\gppoint{gp mark 1}{(3.613,4.119)}
\gppoint{gp mark 1}{(3.651,3.612)}
\gppoint{gp mark 1}{(3.688,2.181)}
\gppoint{gp mark 1}{(3.726,3.951)}
\gppoint{gp mark 1}{(3.764,3.131)}
\gppoint{gp mark 1}{(3.801,3.391)}
\gppoint{gp mark 1}{(3.839,3.560)}
\gppoint{gp mark 1}{(3.877,3.341)}
\gppoint{gp mark 1}{(3.914,2.779)}
\gppoint{gp mark 1}{(3.952,3.243)}
\gppoint{gp mark 1}{(3.990,3.131)}
\gppoint{gp mark 1}{(4.027,3.260)}
\gppoint{gp mark 1}{(4.065,2.960)}
\gppoint{gp mark 1}{(4.103,3.509)}
\gppoint{gp mark 1}{(4.140,3.163)}
\gppoint{gp mark 1}{(4.178,3.441)}
\gppoint{gp mark 1}{(4.216,4.253)}
\gppoint{gp mark 1}{(4.253,3.458)}
\gppoint{gp mark 1}{(4.291,3.084)}
\gppoint{gp mark 1}{(4.329,4.391)}
\gppoint{gp mark 1}{(4.366,3.577)}
\gppoint{gp mark 1}{(4.404,4.081)}
\gppoint{gp mark 1}{(4.441,3.391)}
\gppoint{gp mark 1}{(4.479,4.007)}
\gppoint{gp mark 1}{(4.517,3.037)}
\gppoint{gp mark 1}{(4.554,4.100)}
\gppoint{gp mark 1}{(4.592,4.215)}
\gppoint{gp mark 1}{(4.630,4.312)}
\gppoint{gp mark 1}{(4.667,2.779)}
\gppoint{gp mark 1}{(4.705,3.629)}
\gppoint{gp mark 1}{(4.743,4.532)}
\gppoint{gp mark 1}{(4.780,3.358)}
\gppoint{gp mark 1}{(4.818,3.408)}
\gppoint{gp mark 1}{(4.856,3.988)}
\gppoint{gp mark 1}{(4.893,4.176)}
\gppoint{gp mark 1}{(4.931,4.312)}
\gppoint{gp mark 1}{(4.969,3.969)}
\gppoint{gp mark 1}{(5.006,3.408)}
\gppoint{gp mark 1}{(5.044,3.458)}
\gppoint{gp mark 1}{(5.082,4.215)}
\gppoint{gp mark 1}{(5.119,4.655)}
\gppoint{gp mark 1}{(5.157,3.309)}
\gppoint{gp mark 1}{(5.195,3.969)}
\gppoint{gp mark 1}{(5.232,2.368)}
\gppoint{gp mark 1}{(5.270,4.532)}
\gppoint{gp mark 1}{(5.308,3.116)}
\gppoint{gp mark 1}{(5.345,2.735)}
\gppoint{gp mark 1}{(5.383,2.854)}
\gppoint{gp mark 1}{(5.421,3.629)}
\gppoint{gp mark 1}{(5.458,3.509)}
\gppoint{gp mark 1}{(5.496,3.664)}
\gppoint{gp mark 1}{(5.534,2.677)}
\gppoint{gp mark 1}{(5.571,3.211)}
\gppoint{gp mark 1}{(5.609,4.063)}
\gppoint{gp mark 1}{(5.647,4.273)}
\gppoint{gp mark 1}{(5.684,4.157)}
\gppoint{gp mark 1}{(5.722,3.022)}
\gppoint{gp mark 1}{(5.760,4.273)}
\gppoint{gp mark 1}{(5.797,3.594)}
\gppoint{gp mark 1}{(5.835,3.260)}
\gppoint{gp mark 1}{(5.873,4.273)}
\gppoint{gp mark 1}{(5.910,4.007)}
\gppoint{gp mark 1}{(5.948,2.914)}
\gppoint{gp mark 1}{(5.986,3.878)}
\gppoint{gp mark 1}{(6.023,4.371)}
\gppoint{gp mark 1}{(6.061,4.431)}
\gppoint{gp mark 1}{(6.099,3.699)}
\gppoint{gp mark 1}{(6.136,5.644)}
\gppoint{gp mark 1}{(6.174,4.234)}
\gppoint{gp mark 1}{(6.212,3.375)}
\gppoint{gp mark 1}{(6.249,4.025)}
\gppoint{gp mark 1}{(6.287,3.492)}
\gppoint{gp mark 1}{(6.324,4.739)}
\gppoint{gp mark 1}{(6.362,4.739)}
\gppoint{gp mark 1}{(6.400,4.451)}
\gppoint{gp mark 1}{(6.437,4.676)}
\gppoint{gp mark 1}{(6.475,3.211)}
\gppoint{gp mark 1}{(6.513,3.933)}
\gppoint{gp mark 1}{(6.550,3.752)}
\gppoint{gp mark 1}{(6.588,5.019)}
\gppoint{gp mark 1}{(6.626,3.717)}
\gppoint{gp mark 1}{(6.663,4.292)}
\gppoint{gp mark 1}{(6.701,3.734)}
\gppoint{gp mark 1}{(6.739,3.292)}
\gppoint{gp mark 1}{(6.776,4.371)}
\gppoint{gp mark 1}{(6.814,4.081)}
\gppoint{gp mark 1}{(6.852,5.130)}
\gppoint{gp mark 1}{(6.889,4.572)}
\gppoint{gp mark 1}{(6.927,3.842)}
\gppoint{gp mark 1}{(6.965,4.138)}
\gppoint{gp mark 1}{(7.002,4.655)}
\gppoint{gp mark 1}{(7.040,4.025)}
\gppoint{gp mark 1}{(7.078,4.332)}
\gppoint{gp mark 1}{(7.115,4.471)}
\gppoint{gp mark 1}{(7.153,6.071)}
\gppoint{gp mark 1}{(7.191,3.896)}
\gppoint{gp mark 1}{(7.228,4.676)}
\gppoint{gp mark 1}{(7.266,4.100)}
\gppoint{gp mark 1}{(7.304,2.929)}
\gppoint{gp mark 1}{(7.341,2.534)}
\gppoint{gp mark 1}{(7.379,4.234)}
\gppoint{gp mark 1}{(7.417,4.802)}
\gppoint{gp mark 1}{(7.454,3.824)}
\gppoint{gp mark 1}{(7.492,4.931)}
\gppoint{gp mark 1}{(7.530,3.969)}
\gppoint{gp mark 1}{(7.567,4.593)}
\gppoint{gp mark 1}{(7.605,3.860)}
\gppoint{gp mark 1}{(7.643,5.405)}
\gppoint{gp mark 1}{(7.680,5.220)}
\gppoint{gp mark 1}{(7.718,4.253)}
\gppoint{gp mark 1}{(7.756,3.969)}
\gppoint{gp mark 1}{(7.793,4.195)}
\gppoint{gp mark 1}{(7.831,4.312)}
\gppoint{gp mark 1}{(7.869,4.697)}
\gppoint{gp mark 1}{(7.906,4.371)}
\gppoint{gp mark 1}{(7.944,4.697)}
\gppoint{gp mark 1}{(7.982,2.750)}
\gppoint{gp mark 1}{(8.019,4.655)}
\gppoint{gp mark 1}{(8.057,5.358)}
\gppoint{gp mark 1}{(8.094,4.718)}
\gppoint{gp mark 1}{(8.132,4.063)}
\gppoint{gp mark 1}{(8.170,5.019)}
\gppoint{gp mark 1}{(8.207,5.041)}
\gppoint{gp mark 1}{(8.245,4.823)}
\gppoint{gp mark 1}{(8.283,5.620)}
\gppoint{gp mark 1}{(8.320,4.119)}
\gppoint{gp mark 1}{(8.358,4.195)}
\gppoint{gp mark 1}{(8.396,4.552)}
\gppoint{gp mark 1}{(8.433,3.788)}
\gppoint{gp mark 1}{(8.471,4.888)}
\gppoint{gp mark 1}{(8.509,4.215)}
\gppoint{gp mark 1}{(8.546,4.431)}
\gppoint{gp mark 1}{(8.584,5.595)}
\gppoint{gp mark 1}{(8.622,4.025)}
\gppoint{gp mark 1}{(8.659,4.312)}
\gppoint{gp mark 1}{(8.697,5.198)}
\gppoint{gp mark 1}{(8.735,3.951)}
\gppoint{gp mark 1}{(8.772,4.411)}
\gppoint{gp mark 1}{(8.810,5.063)}
\gppoint{gp mark 1}{(8.848,4.823)}
\gppoint{gp mark 1}{(8.885,5.019)}
\gppoint{gp mark 1}{(8.923,4.760)}
\gppoint{gp mark 1}{(8.961,5.107)}
\gppoint{gp mark 1}{(8.998,4.044)}
\gppoint{gp mark 1}{(9.036,6.176)}
\gppoint{gp mark 1}{(9.074,4.431)}
\gppoint{gp mark 1}{(9.111,5.644)}
\gppoint{gp mark 1}{(9.149,4.655)}
\gppoint{gp mark 1}{(9.187,6.097)}
\gppoint{gp mark 1}{(9.224,5.019)}
\gppoint{gp mark 1}{(9.262,4.866)}
\gppoint{gp mark 1}{(9.300,5.243)}
\gppoint{gp mark 1}{(9.337,4.845)}
\gppoint{gp mark 1}{(9.375,4.157)}
\gppoint{gp mark 1}{(9.413,5.085)}
\gppoint{gp mark 1}{(9.450,4.431)}
\gppoint{gp mark 1}{(9.488,4.614)}
\gppoint{gp mark 1}{(9.526,5.841)}
\gppoint{gp mark 1}{(9.563,5.335)}
\gppoint{gp mark 1}{(9.601,6.255)}
\gppoint{gp mark 1}{(9.639,5.358)}
\gppoint{gp mark 1}{(9.676,5.917)}
\gppoint{gp mark 1}{(9.714,6.019)}
\gppoint{gp mark 1}{(9.752,4.676)}
\gppoint{gp mark 1}{(9.789,3.543)}
\gppoint{gp mark 1}{(9.827,4.593)}
\gppoint{gp mark 1}{(9.865,6.926)}
\gppoint{gp mark 1}{(9.902,5.917)}
\gppoint{gp mark 1}{(9.940,6.202)}
\gppoint{gp mark 1}{(9.978,6.810)}
\gppoint{gp mark 1}{(10.015,4.081)}
\gppoint{gp mark 1}{(10.053,5.107)}
\gppoint{gp mark 1}{(10.090,5.994)}
\gppoint{gp mark 1}{(10.128,4.781)}
\gppoint{gp mark 1}{(10.166,5.841)}
\gppoint{gp mark 1}{(10.203,5.405)}
\gppoint{gp mark 1}{(10.241,5.547)}
\gppoint{gp mark 1}{(10.279,4.044)}
\gppoint{gp mark 1}{(10.316,6.019)}
\gppoint{gp mark 1}{(10.354,4.552)}
\gppoint{gp mark 1}{(10.392,5.243)}
\gppoint{gp mark 1}{(10.429,4.823)}
\gppoint{gp mark 1}{(10.467,4.511)}
\gppoint{gp mark 1}{(10.505,6.045)}
\gppoint{gp mark 1}{(10.542,4.451)}
\gppoint{gp mark 1}{(10.580,4.697)}
\gppoint{gp mark 1}{(10.618,4.866)}
\gppoint{gp mark 1}{(10.655,5.041)}
\gppoint{gp mark 1}{(10.693,4.044)}
\gppoint{gp mark 1}{(10.731,4.572)}
\gppoint{gp mark 1}{(10.093,1.319)}
\gpcolor{color=gp lt color border}
\draw[gp path] (1.504,7.251)--(1.504,0.985)--(10.919,0.985)--(10.919,7.251)--cycle;
\gpdefrectangularnode{gp plot 1}{\pgfpoint{1.504cm}{0.985cm}}{\pgfpoint{10.919cm}{7.251cm}}
\end{tikzpicture}
}
    \subcaption{Program size not controlled.}\label{fig:rand-no-limit}
  \end{minipage}
  \begin{minipage}[b]{1.0\linewidth}
    \adjustbox{width=1.0\textwidth}{
\begin{tikzpicture}[gnuplot]
\path (0.000,0.000) rectangle (12.700,7.620);
\gpcolor{color=gp lt color border}
\gpsetlinetype{gp lt border}
\gpsetlinewidth{1.00}
\draw[gp path] (1.504,0.985)--(1.684,0.985);
\node[gp node right] at (1.320,0.985) { 172};
\draw[gp path] (1.504,3.074)--(1.684,3.074);
\node[gp node right] at (1.320,3.074) { 173};
\draw[gp path] (1.504,5.162)--(1.684,5.162);
\node[gp node right] at (1.320,5.162) { 174};
\draw[gp path] (1.504,7.251)--(1.684,7.251);
\node[gp node right] at (1.320,7.251) { 175};
\draw[gp path] (1.504,0.985)--(1.504,1.165);
\draw[gp path] (1.504,7.251)--(1.504,7.071);
\node[gp node center] at (1.504,0.677) { 0};
\draw[gp path] (3.387,0.985)--(3.387,1.165);
\draw[gp path] (3.387,7.251)--(3.387,7.071);
\node[gp node center] at (3.387,0.677) { 50};
\draw[gp path] (5.270,0.985)--(5.270,1.165);
\draw[gp path] (5.270,7.251)--(5.270,7.071);
\node[gp node center] at (5.270,0.677) { 100};
\draw[gp path] (7.153,0.985)--(7.153,1.165);
\draw[gp path] (7.153,7.251)--(7.153,7.071);
\node[gp node center] at (7.153,0.677) { 150};
\draw[gp path] (9.036,0.985)--(9.036,1.165);
\draw[gp path] (9.036,7.251)--(9.036,7.071);
\node[gp node center] at (9.036,0.677) { 200};
\draw[gp path] (10.919,0.985)--(10.919,1.165);
\draw[gp path] (10.919,7.251)--(10.919,7.071);
\node[gp node center] at (10.919,0.677) { 250};
\draw[gp path] (10.919,0.985)--(10.739,0.985);
\node[gp node left] at (11.103,0.985) { 4};
\draw[gp path] (10.919,1.768)--(10.739,1.768);
\node[gp node left] at (11.103,1.768) { 6};
\draw[gp path] (10.919,2.552)--(10.739,2.552);
\node[gp node left] at (11.103,2.552) { 8};
\draw[gp path] (10.919,3.335)--(10.739,3.335);
\node[gp node left] at (11.103,3.335) { 10};
\draw[gp path] (10.919,4.118)--(10.739,4.118);
\node[gp node left] at (11.103,4.118) { 12};
\draw[gp path] (10.919,4.901)--(10.739,4.901);
\node[gp node left] at (11.103,4.901) { 14};
\draw[gp path] (10.919,5.685)--(10.739,5.685);
\node[gp node left] at (11.103,5.685) { 16};
\draw[gp path] (10.919,6.468)--(10.739,6.468);
\node[gp node left] at (11.103,6.468) { 18};
\draw[gp path] (10.919,7.251)--(10.739,7.251);
\node[gp node left] at (11.103,7.251) { 20};
\draw[gp path] (1.504,7.251)--(1.504,0.985)--(10.919,0.985)--(10.919,7.251)--cycle;
\node[gp node center,rotate=-270] at (0.246,4.118) {Avg. LOC};
\node[gp node center,rotate=-270] at (11.992,4.118) {\% Neutral Variants};
\node[gp node center] at (6.211,0.215) {Number of Applied Mutations};
\node[gp node right] at (9.451,1.627) {Avg. LOC};
\gpcolor{color=gp lt color 7}
\gpsetpointsize{4.00}
\gppoint{gp mark 8}{(1.542,5.806)}
\gppoint{gp mark 8}{(1.579,6.119)}
\gppoint{gp mark 8}{(1.617,5.131)}
\gppoint{gp mark 8}{(1.655,4.953)}
\gppoint{gp mark 8}{(1.692,4.072)}
\gppoint{gp mark 8}{(1.730,4.051)}
\gppoint{gp mark 8}{(1.768,4.009)}
\gppoint{gp mark 8}{(1.805,3.748)}
\gppoint{gp mark 8}{(1.843,3.790)}
\gppoint{gp mark 8}{(1.881,3.583)}
\gppoint{gp mark 8}{(1.918,3.738)}
\gppoint{gp mark 8}{(1.956,3.698)}
\gppoint{gp mark 8}{(1.994,3.780)}
\gppoint{gp mark 8}{(2.031,3.562)}
\gppoint{gp mark 8}{(2.069,3.874)}
\gppoint{gp mark 8}{(2.107,3.811)}
\gppoint{gp mark 8}{(2.144,3.790)}
\gppoint{gp mark 8}{(2.182,3.769)}
\gppoint{gp mark 8}{(2.220,3.895)}
\gppoint{gp mark 8}{(2.257,3.688)}
\gppoint{gp mark 8}{(2.295,4.041)}
\gppoint{gp mark 8}{(2.333,3.667)}
\gppoint{gp mark 8}{(2.370,3.395)}
\gppoint{gp mark 8}{(2.408,3.343)}
\gppoint{gp mark 8}{(2.446,3.042)}
\gppoint{gp mark 8}{(2.483,3.262)}
\gppoint{gp mark 8}{(2.521,3.136)}
\gppoint{gp mark 8}{(2.558,2.721)}
\gppoint{gp mark 8}{(2.596,3.283)}
\gppoint{gp mark 8}{(2.634,2.886)}
\gppoint{gp mark 8}{(2.671,3.168)}
\gppoint{gp mark 8}{(2.709,2.846)}
\gppoint{gp mark 8}{(2.747,2.679)}
\gppoint{gp mark 8}{(2.784,2.679)}
\gppoint{gp mark 8}{(2.822,2.554)}
\gppoint{gp mark 8}{(2.860,2.907)}
\gppoint{gp mark 8}{(2.897,3.147)}
\gppoint{gp mark 8}{(2.935,3.063)}
\gppoint{gp mark 8}{(2.973,3.032)}
\gppoint{gp mark 8}{(3.010,3.283)}
\gppoint{gp mark 8}{(3.048,3.448)}
\gppoint{gp mark 8}{(3.086,3.542)}
\gppoint{gp mark 8}{(3.123,3.698)}
\gppoint{gp mark 8}{(3.161,4.051)}
\gppoint{gp mark 8}{(3.199,3.895)}
\gppoint{gp mark 8}{(3.236,4.009)}
\gppoint{gp mark 8}{(3.274,4.143)}
\gppoint{gp mark 8}{(3.312,4.300)}
\gppoint{gp mark 8}{(3.349,4.394)}
\gppoint{gp mark 8}{(3.387,4.580)}
\gppoint{gp mark 8}{(3.425,4.642)}
\gppoint{gp mark 8}{(3.462,4.425)}
\gppoint{gp mark 8}{(3.500,4.694)}
\gppoint{gp mark 8}{(3.538,4.509)}
\gppoint{gp mark 8}{(3.575,4.621)}
\gppoint{gp mark 8}{(3.613,4.611)}
\gppoint{gp mark 8}{(3.651,4.757)}
\gppoint{gp mark 8}{(3.688,4.642)}
\gppoint{gp mark 8}{(3.726,4.279)}
\gppoint{gp mark 8}{(3.764,3.957)}
\gppoint{gp mark 8}{(3.801,4.030)}
\gppoint{gp mark 8}{(3.839,4.143)}
\gppoint{gp mark 8}{(3.877,4.020)}
\gppoint{gp mark 8}{(3.914,4.030)}
\gppoint{gp mark 8}{(3.952,4.477)}
\gppoint{gp mark 8}{(3.990,4.446)}
\gppoint{gp mark 8}{(4.027,4.611)}
\gppoint{gp mark 8}{(4.065,4.736)}
\gppoint{gp mark 8}{(4.103,4.841)}
\gppoint{gp mark 8}{(4.140,4.914)}
\gppoint{gp mark 8}{(4.178,5.110)}
\gppoint{gp mark 8}{(4.216,5.173)}
\gppoint{gp mark 8}{(4.253,4.872)}
\gppoint{gp mark 8}{(4.291,4.809)}
\gppoint{gp mark 8}{(4.329,5.027)}
\gppoint{gp mark 8}{(4.366,4.830)}
\gppoint{gp mark 8}{(4.404,4.924)}
\gppoint{gp mark 8}{(4.441,4.694)}
\gppoint{gp mark 8}{(4.479,4.778)}
\gppoint{gp mark 8}{(4.517,4.632)}
\gppoint{gp mark 8}{(4.554,4.872)}
\gppoint{gp mark 8}{(4.592,4.953)}
\gppoint{gp mark 8}{(4.630,4.914)}
\gppoint{gp mark 8}{(4.667,5.141)}
\gppoint{gp mark 8}{(4.705,5.235)}
\gppoint{gp mark 8}{(4.743,5.267)}
\gppoint{gp mark 8}{(4.780,5.474)}
\gppoint{gp mark 8}{(4.818,5.609)}
\gppoint{gp mark 8}{(4.856,5.400)}
\gppoint{gp mark 8}{(4.893,5.526)}
\gppoint{gp mark 8}{(4.931,5.776)}
\gppoint{gp mark 8}{(4.969,5.900)}
\gppoint{gp mark 8}{(5.006,5.921)}
\gppoint{gp mark 8}{(5.044,5.745)}
\gppoint{gp mark 8}{(5.082,6.067)}
\gppoint{gp mark 8}{(5.119,5.931)}
\gppoint{gp mark 8}{(5.157,5.952)}
\gppoint{gp mark 8}{(5.195,6.025)}
\gppoint{gp mark 8}{(5.232,6.232)}
\gppoint{gp mark 8}{(5.270,6.108)}
\gppoint{gp mark 8}{(5.308,6.140)}
\gppoint{gp mark 8}{(5.345,5.962)}
\gppoint{gp mark 8}{(5.383,6.067)}
\gppoint{gp mark 8}{(5.421,6.004)}
\gppoint{gp mark 8}{(5.458,6.119)}
\gppoint{gp mark 8}{(5.496,5.921)}
\gppoint{gp mark 8}{(5.534,5.837)}
\gppoint{gp mark 8}{(5.571,5.931)}
\gppoint{gp mark 8}{(5.609,5.931)}
\gppoint{gp mark 8}{(5.647,5.868)}
\gppoint{gp mark 8}{(5.684,5.776)}
\gppoint{gp mark 8}{(5.722,5.588)}
\gppoint{gp mark 8}{(5.760,5.662)}
\gppoint{gp mark 8}{(5.797,5.682)}
\gppoint{gp mark 8}{(5.835,5.703)}
\gppoint{gp mark 8}{(5.873,5.547)}
\gppoint{gp mark 8}{(5.910,5.505)}
\gppoint{gp mark 8}{(5.948,5.568)}
\gppoint{gp mark 8}{(5.986,5.672)}
\gppoint{gp mark 8}{(6.023,5.620)}
\gppoint{gp mark 8}{(6.061,5.641)}
\gppoint{gp mark 8}{(6.099,5.557)}
\gppoint{gp mark 8}{(6.136,5.827)}
\gppoint{gp mark 8}{(6.174,5.910)}
\gppoint{gp mark 8}{(6.212,5.910)}
\gppoint{gp mark 8}{(6.249,5.931)}
\gppoint{gp mark 8}{(6.287,5.910)}
\gppoint{gp mark 8}{(6.324,5.962)}
\gppoint{gp mark 8}{(6.362,5.787)}
\gppoint{gp mark 8}{(6.400,5.879)}
\gppoint{gp mark 8}{(6.437,6.015)}
\gppoint{gp mark 8}{(6.475,5.806)}
\gppoint{gp mark 8}{(6.513,6.067)}
\gppoint{gp mark 8}{(6.550,5.994)}
\gppoint{gp mark 8}{(6.588,6.025)}
\gppoint{gp mark 8}{(6.626,6.025)}
\gppoint{gp mark 8}{(6.663,6.025)}
\gppoint{gp mark 8}{(6.701,6.004)}
\gppoint{gp mark 8}{(6.739,5.973)}
\gppoint{gp mark 8}{(6.776,6.025)}
\gppoint{gp mark 8}{(6.814,5.941)}
\gppoint{gp mark 8}{(6.852,6.129)}
\gppoint{gp mark 8}{(6.889,6.482)}
\gppoint{gp mark 8}{(6.927,6.493)}
\gppoint{gp mark 8}{(6.965,6.211)}
\gppoint{gp mark 8}{(7.002,6.211)}
\gppoint{gp mark 8}{(7.040,6.067)}
\gppoint{gp mark 8}{(7.078,5.756)}
\gppoint{gp mark 8}{(7.115,5.693)}
\gppoint{gp mark 8}{(7.153,5.931)}
\gppoint{gp mark 8}{(7.191,5.994)}
\gppoint{gp mark 8}{(7.228,6.150)}
\gppoint{gp mark 8}{(7.266,6.129)}
\gppoint{gp mark 8}{(7.304,6.067)}
\gppoint{gp mark 8}{(7.341,6.015)}
\gppoint{gp mark 8}{(7.379,5.962)}
\gppoint{gp mark 8}{(7.417,5.994)}
\gppoint{gp mark 8}{(7.454,5.910)}
\gppoint{gp mark 8}{(7.492,5.693)}
\gppoint{gp mark 8}{(7.530,5.766)}
\gppoint{gp mark 8}{(7.567,5.795)}
\gppoint{gp mark 8}{(7.605,5.484)}
\gppoint{gp mark 8}{(7.643,5.183)}
\gppoint{gp mark 8}{(7.680,4.788)}
\gppoint{gp mark 8}{(7.718,5.079)}
\gppoint{gp mark 8}{(7.756,5.215)}
\gppoint{gp mark 8}{(7.793,5.027)}
\gppoint{gp mark 8}{(7.831,5.131)}
\gppoint{gp mark 8}{(7.869,5.079)}
\gppoint{gp mark 8}{(7.906,5.329)}
\gppoint{gp mark 8}{(7.944,5.526)}
\gppoint{gp mark 8}{(7.982,5.474)}
\gppoint{gp mark 8}{(8.019,5.235)}
\gppoint{gp mark 8}{(8.057,4.985)}
\gppoint{gp mark 8}{(8.094,5.047)}
\gppoint{gp mark 8}{(8.132,5.089)}
\gppoint{gp mark 8}{(8.170,5.037)}
\gppoint{gp mark 8}{(8.207,4.914)}
\gppoint{gp mark 8}{(8.245,5.037)}
\gppoint{gp mark 8}{(8.283,5.173)}
\gppoint{gp mark 8}{(8.320,5.089)}
\gppoint{gp mark 8}{(8.358,4.935)}
\gppoint{gp mark 8}{(8.396,5.100)}
\gppoint{gp mark 8}{(8.433,5.037)}
\gppoint{gp mark 8}{(8.471,5.037)}
\gppoint{gp mark 8}{(8.509,5.121)}
\gppoint{gp mark 8}{(8.546,5.568)}
\gppoint{gp mark 8}{(8.584,5.745)}
\gppoint{gp mark 8}{(8.622,5.962)}
\gppoint{gp mark 8}{(8.659,6.088)}
\gppoint{gp mark 8}{(8.697,6.202)}
\gppoint{gp mark 8}{(8.735,6.108)}
\gppoint{gp mark 8}{(8.772,6.119)}
\gppoint{gp mark 8}{(8.810,6.056)}
\gppoint{gp mark 8}{(8.848,5.994)}
\gppoint{gp mark 8}{(8.885,6.025)}
\gppoint{gp mark 8}{(8.923,6.119)}
\gppoint{gp mark 8}{(8.961,6.202)}
\gppoint{gp mark 8}{(8.998,6.441)}
\gppoint{gp mark 8}{(9.036,6.409)}
\gppoint{gp mark 8}{(9.074,6.555)}
\gppoint{gp mark 8}{(9.111,6.566)}
\gppoint{gp mark 8}{(9.149,6.493)}
\gppoint{gp mark 8}{(9.187,6.503)}
\gppoint{gp mark 8}{(9.224,6.451)}
\gppoint{gp mark 8}{(9.262,6.367)}
\gppoint{gp mark 8}{(9.300,6.493)}
\gppoint{gp mark 8}{(9.337,6.555)}
\gppoint{gp mark 8}{(9.375,6.700)}
\gppoint{gp mark 8}{(9.413,6.940)}
\gppoint{gp mark 8}{(9.450,6.919)}
\gppoint{gp mark 8}{(9.488,6.940)}
\gppoint{gp mark 8}{(9.526,6.783)}
\gppoint{gp mark 8}{(9.563,6.762)}
\gppoint{gp mark 8}{(9.601,6.731)}
\gppoint{gp mark 8}{(9.639,6.689)}
\gppoint{gp mark 8}{(9.676,6.762)}
\gppoint{gp mark 8}{(9.714,6.618)}
\gppoint{gp mark 8}{(9.752,6.647)}
\gppoint{gp mark 8}{(9.789,6.576)}
\gppoint{gp mark 8}{(9.827,6.545)}
\gppoint{gp mark 8}{(9.865,6.535)}
\gppoint{gp mark 8}{(9.902,6.626)}
\gppoint{gp mark 8}{(9.940,6.668)}
\gppoint{gp mark 8}{(9.978,6.814)}
\gppoint{gp mark 8}{(10.015,6.679)}
\gppoint{gp mark 8}{(10.053,6.658)}
\gppoint{gp mark 8}{(10.090,6.524)}
\gppoint{gp mark 8}{(10.128,6.720)}
\gppoint{gp mark 8}{(10.166,6.658)}
\gppoint{gp mark 8}{(10.203,6.647)}
\gppoint{gp mark 8}{(10.241,6.626)}
\gppoint{gp mark 8}{(10.279,6.637)}
\gppoint{gp mark 8}{(10.316,6.658)}
\gppoint{gp mark 8}{(10.354,6.668)}
\gppoint{gp mark 8}{(10.392,6.647)}
\gppoint{gp mark 8}{(10.429,6.647)}
\gppoint{gp mark 8}{(10.467,6.679)}
\gppoint{gp mark 8}{(10.505,6.814)}
\gppoint{gp mark 8}{(10.542,6.908)}
\gppoint{gp mark 8}{(10.580,6.898)}
\gppoint{gp mark 8}{(10.618,6.940)}
\gppoint{gp mark 8}{(10.655,6.950)}
\gppoint{gp mark 8}{(10.693,7.105)}
\gppoint{gp mark 8}{(10.731,7.034)}
\gppoint{gp mark 8}{(10.768,6.908)}
\gppoint{gp mark 8}{(10.806,6.835)}
\gppoint{gp mark 8}{(10.844,6.888)}
\gppoint{gp mark 8}{(10.881,6.867)}
\gppoint{gp mark 8}{(10.919,6.741)}
\gppoint{gp mark 8}{(10.093,1.627)}
\gpcolor{color=gp lt color border}
\node[gp node right] at (9.451,1.319) {\% Neutral Variants};
\gpcolor{rgb color={0.000,0.000,1.000}}
\gppoint{gp mark 1}{(1.542,1.703)}
\gppoint{gp mark 1}{(1.579,4.911)}
\gppoint{gp mark 1}{(1.617,3.460)}
\gppoint{gp mark 1}{(1.655,4.739)}
\gppoint{gp mark 1}{(1.692,4.675)}
\gppoint{gp mark 1}{(1.730,4.551)}
\gppoint{gp mark 1}{(1.768,5.102)}
\gppoint{gp mark 1}{(1.805,4.498)}
\gppoint{gp mark 1}{(1.843,5.425)}
\gppoint{gp mark 1}{(1.881,4.578)}
\gppoint{gp mark 1}{(1.918,4.518)}
\gppoint{gp mark 1}{(1.956,5.078)}
\gppoint{gp mark 1}{(1.994,4.302)}
\gppoint{gp mark 1}{(2.031,4.171)}
\gppoint{gp mark 1}{(2.069,4.747)}
\gppoint{gp mark 1}{(2.107,4.114)}
\gppoint{gp mark 1}{(2.144,4.647)}
\gppoint{gp mark 1}{(2.182,4.934)}
\gppoint{gp mark 1}{(2.220,5.062)}
\gppoint{gp mark 1}{(2.257,4.966)}
\gppoint{gp mark 1}{(2.295,5.195)}
\gppoint{gp mark 1}{(2.333,4.696)}
\gppoint{gp mark 1}{(2.370,4.682)}
\gppoint{gp mark 1}{(2.408,5.229)}
\gppoint{gp mark 1}{(2.446,4.865)}
\gppoint{gp mark 1}{(2.483,5.094)}
\gppoint{gp mark 1}{(2.521,4.850)}
\gppoint{gp mark 1}{(2.558,5.070)}
\gppoint{gp mark 1}{(2.596,4.896)}
\gppoint{gp mark 1}{(2.634,5.111)}
\gppoint{gp mark 1}{(2.671,5.246)}
\gppoint{gp mark 1}{(2.709,5.547)}
\gppoint{gp mark 1}{(2.747,5.144)}
\gppoint{gp mark 1}{(2.784,5.037)}
\gppoint{gp mark 1}{(2.822,5.078)}
\gppoint{gp mark 1}{(2.860,5.229)}
\gppoint{gp mark 1}{(2.897,4.981)}
\gppoint{gp mark 1}{(2.935,4.718)}
\gppoint{gp mark 1}{(2.973,5.462)}
\gppoint{gp mark 1}{(3.010,5.471)}
\gppoint{gp mark 1}{(3.048,4.711)}
\gppoint{gp mark 1}{(3.086,5.062)}
\gppoint{gp mark 1}{(3.123,5.144)}
\gppoint{gp mark 1}{(3.161,4.619)}
\gppoint{gp mark 1}{(3.199,5.220)}
\gppoint{gp mark 1}{(3.236,4.873)}
\gppoint{gp mark 1}{(3.274,5.625)}
\gppoint{gp mark 1}{(3.312,5.062)}
\gppoint{gp mark 1}{(3.349,4.485)}
\gppoint{gp mark 1}{(3.387,4.414)}
\gppoint{gp mark 1}{(3.425,4.465)}
\gppoint{gp mark 1}{(3.462,4.640)}
\gppoint{gp mark 1}{(3.500,4.459)}
\gppoint{gp mark 1}{(3.538,5.352)}
\gppoint{gp mark 1}{(3.575,4.725)}
\gppoint{gp mark 1}{(3.613,4.865)}
\gppoint{gp mark 1}{(3.651,5.102)}
\gppoint{gp mark 1}{(3.688,5.272)}
\gppoint{gp mark 1}{(3.726,4.578)}
\gppoint{gp mark 1}{(3.764,5.308)}
\gppoint{gp mark 1}{(3.801,4.835)}
\gppoint{gp mark 1}{(3.839,4.538)}
\gppoint{gp mark 1}{(3.877,5.029)}
\gppoint{gp mark 1}{(3.914,4.696)}
\gppoint{gp mark 1}{(3.952,5.136)}
\gppoint{gp mark 1}{(3.990,5.029)}
\gppoint{gp mark 1}{(4.027,5.062)}
\gppoint{gp mark 1}{(4.065,5.152)}
\gppoint{gp mark 1}{(4.103,4.843)}
\gppoint{gp mark 1}{(4.140,5.136)}
\gppoint{gp mark 1}{(4.178,4.704)}
\gppoint{gp mark 1}{(4.216,4.835)}
\gppoint{gp mark 1}{(4.253,4.395)}
\gppoint{gp mark 1}{(4.291,4.754)}
\gppoint{gp mark 1}{(4.329,4.934)}
\gppoint{gp mark 1}{(4.366,5.094)}
\gppoint{gp mark 1}{(4.404,4.452)}
\gppoint{gp mark 1}{(4.441,5.029)}
\gppoint{gp mark 1}{(4.479,4.820)}
\gppoint{gp mark 1}{(4.517,5.325)}
\gppoint{gp mark 1}{(4.554,6.124)}
\gppoint{gp mark 1}{(4.592,4.903)}
\gppoint{gp mark 1}{(4.630,5.212)}
\gppoint{gp mark 1}{(4.667,5.490)}
\gppoint{gp mark 1}{(4.705,5.053)}
\gppoint{gp mark 1}{(4.743,5.178)}
\gppoint{gp mark 1}{(4.780,5.538)}
\gppoint{gp mark 1}{(4.818,5.094)}
\gppoint{gp mark 1}{(4.856,5.111)}
\gppoint{gp mark 1}{(4.893,5.557)}
\gppoint{gp mark 1}{(4.931,4.873)}
\gppoint{gp mark 1}{(4.969,5.246)}
\gppoint{gp mark 1}{(5.006,5.264)}
\gppoint{gp mark 1}{(5.044,4.654)}
\gppoint{gp mark 1}{(5.082,4.880)}
\gppoint{gp mark 1}{(5.119,4.783)}
\gppoint{gp mark 1}{(5.157,5.238)}
\gppoint{gp mark 1}{(5.195,4.571)}
\gppoint{gp mark 1}{(5.232,4.654)}
\gppoint{gp mark 1}{(5.270,4.865)}
\gppoint{gp mark 1}{(5.308,5.161)}
\gppoint{gp mark 1}{(5.345,4.478)}
\gppoint{gp mark 1}{(5.383,5.343)}
\gppoint{gp mark 1}{(5.421,4.551)}
\gppoint{gp mark 1}{(5.458,5.264)}
\gppoint{gp mark 1}{(5.496,5.655)}
\gppoint{gp mark 1}{(5.534,5.316)}
\gppoint{gp mark 1}{(5.571,5.625)}
\gppoint{gp mark 1}{(5.609,5.086)}
\gppoint{gp mark 1}{(5.647,4.880)}
\gppoint{gp mark 1}{(5.684,5.013)}
\gppoint{gp mark 1}{(5.722,5.509)}
\gppoint{gp mark 1}{(5.760,4.578)}
\gppoint{gp mark 1}{(5.797,5.586)}
\gppoint{gp mark 1}{(5.835,5.388)}
\gppoint{gp mark 1}{(5.873,5.615)}
\gppoint{gp mark 1}{(5.910,4.661)}
\gppoint{gp mark 1}{(5.948,5.745)}
\gppoint{gp mark 1}{(5.986,4.776)}
\gppoint{gp mark 1}{(6.023,4.927)}
\gppoint{gp mark 1}{(6.061,5.596)}
\gppoint{gp mark 1}{(6.099,4.619)}
\gppoint{gp mark 1}{(6.136,5.674)}
\gppoint{gp mark 1}{(6.174,5.325)}
\gppoint{gp mark 1}{(6.212,5.398)}
\gppoint{gp mark 1}{(6.249,5.186)}
\gppoint{gp mark 1}{(6.287,5.860)}
\gppoint{gp mark 1}{(6.324,4.934)}
\gppoint{gp mark 1}{(6.362,5.062)}
\gppoint{gp mark 1}{(6.400,5.037)}
\gppoint{gp mark 1}{(6.437,5.490)}
\gppoint{gp mark 1}{(6.475,5.178)}
\gppoint{gp mark 1}{(6.513,5.625)}
\gppoint{gp mark 1}{(6.550,5.078)}
\gppoint{gp mark 1}{(6.588,4.558)}
\gppoint{gp mark 1}{(6.626,5.220)}
\gppoint{gp mark 1}{(6.663,4.919)}
\gppoint{gp mark 1}{(6.701,4.689)}
\gppoint{gp mark 1}{(6.739,5.195)}
\gppoint{gp mark 1}{(6.776,5.576)}
\gppoint{gp mark 1}{(6.814,5.195)}
\gppoint{gp mark 1}{(6.852,5.053)}
\gppoint{gp mark 1}{(6.889,5.127)}
\gppoint{gp mark 1}{(6.927,5.264)}
\gppoint{gp mark 1}{(6.965,5.290)}
\gppoint{gp mark 1}{(7.002,4.934)}
\gppoint{gp mark 1}{(7.040,4.942)}
\gppoint{gp mark 1}{(7.078,5.715)}
\gppoint{gp mark 1}{(7.115,5.766)}
\gppoint{gp mark 1}{(7.153,5.005)}
\gppoint{gp mark 1}{(7.191,5.136)}
\gppoint{gp mark 1}{(7.228,5.665)}
\gppoint{gp mark 1}{(7.266,5.290)}
\gppoint{gp mark 1}{(7.304,5.102)}
\gppoint{gp mark 1}{(7.341,5.655)}
\gppoint{gp mark 1}{(7.379,5.264)}
\gppoint{gp mark 1}{(7.417,6.067)}
\gppoint{gp mark 1}{(7.454,5.635)}
\gppoint{gp mark 1}{(7.492,5.547)}
\gppoint{gp mark 1}{(7.530,6.090)}
\gppoint{gp mark 1}{(7.567,4.942)}
\gppoint{gp mark 1}{(7.605,5.334)}
\gppoint{gp mark 1}{(7.643,5.361)}
\gppoint{gp mark 1}{(7.680,5.500)}
\gppoint{gp mark 1}{(7.718,5.334)}
\gppoint{gp mark 1}{(7.756,5.557)}
\gppoint{gp mark 1}{(7.793,6.147)}
\gppoint{gp mark 1}{(7.831,5.735)}
\gppoint{gp mark 1}{(7.869,5.586)}
\gppoint{gp mark 1}{(7.906,5.735)}
\gppoint{gp mark 1}{(7.944,5.576)}
\gppoint{gp mark 1}{(7.982,4.958)}
\gppoint{gp mark 1}{(8.019,5.715)}
\gppoint{gp mark 1}{(8.057,6.387)}
\gppoint{gp mark 1}{(8.094,5.665)}
\gppoint{gp mark 1}{(8.132,5.566)}
\gppoint{gp mark 1}{(8.170,6.462)}
\gppoint{gp mark 1}{(8.207,5.913)}
\gppoint{gp mark 1}{(8.245,6.253)}
\gppoint{gp mark 1}{(8.283,6.449)}
\gppoint{gp mark 1}{(8.320,6.424)}
\gppoint{gp mark 1}{(8.358,6.301)}
\gppoint{gp mark 1}{(8.396,6.067)}
\gppoint{gp mark 1}{(8.433,6.147)}
\gppoint{gp mark 1}{(8.471,5.596)}
\gppoint{gp mark 1}{(8.509,5.870)}
\gppoint{gp mark 1}{(8.546,5.715)}
\gppoint{gp mark 1}{(8.584,5.892)}
\gppoint{gp mark 1}{(8.622,6.218)}
\gppoint{gp mark 1}{(8.659,6.206)}
\gppoint{gp mark 1}{(8.697,5.870)}
\gppoint{gp mark 1}{(8.735,5.308)}
\gppoint{gp mark 1}{(8.772,5.557)}
\gppoint{gp mark 1}{(8.810,5.136)}
\gppoint{gp mark 1}{(8.848,5.490)}
\gppoint{gp mark 1}{(8.885,6.023)}
\gppoint{gp mark 1}{(8.923,6.012)}
\gppoint{gp mark 1}{(8.961,5.471)}
\gppoint{gp mark 1}{(8.998,5.576)}
\gppoint{gp mark 1}{(9.036,5.528)}
\gppoint{gp mark 1}{(9.074,5.745)}
\gppoint{gp mark 1}{(9.111,5.766)}
\gppoint{gp mark 1}{(9.149,4.981)}
\gppoint{gp mark 1}{(9.187,5.037)}
\gppoint{gp mark 1}{(9.224,5.352)}
\gppoint{gp mark 1}{(9.262,5.425)}
\gppoint{gp mark 1}{(9.300,6.182)}
\gppoint{gp mark 1}{(9.337,6.241)}
\gppoint{gp mark 1}{(9.375,6.375)}
\gppoint{gp mark 1}{(9.413,5.892)}
\gppoint{gp mark 1}{(9.450,5.195)}
\gppoint{gp mark 1}{(9.488,5.547)}
\gppoint{gp mark 1}{(9.526,6.124)}
\gppoint{gp mark 1}{(9.563,5.735)}
\gppoint{gp mark 1}{(9.601,6.000)}
\gppoint{gp mark 1}{(9.639,5.481)}
\gppoint{gp mark 1}{(9.676,6.067)}
\gppoint{gp mark 1}{(9.714,5.635)}
\gppoint{gp mark 1}{(9.752,5.229)}
\gppoint{gp mark 1}{(9.789,5.924)}
\gppoint{gp mark 1}{(9.827,6.090)}
\gppoint{gp mark 1}{(9.865,6.253)}
\gppoint{gp mark 1}{(9.902,5.797)}
\gppoint{gp mark 1}{(9.940,5.144)}
\gppoint{gp mark 1}{(9.978,5.361)}
\gppoint{gp mark 1}{(10.015,5.870)}
\gppoint{gp mark 1}{(10.053,5.860)}
\gppoint{gp mark 1}{(10.090,6.000)}
\gppoint{gp mark 1}{(10.128,5.776)}
\gppoint{gp mark 1}{(10.166,6.147)}
\gppoint{gp mark 1}{(10.203,6.362)}
\gppoint{gp mark 1}{(10.241,5.379)}
\gppoint{gp mark 1}{(10.279,6.301)}
\gppoint{gp mark 1}{(10.316,5.705)}
\gppoint{gp mark 1}{(10.354,6.412)}
\gppoint{gp mark 1}{(10.392,5.471)}
\gppoint{gp mark 1}{(10.429,5.935)}
\gppoint{gp mark 1}{(10.467,4.989)}
\gppoint{gp mark 1}{(10.505,5.334)}
\gppoint{gp mark 1}{(10.542,5.528)}
\gppoint{gp mark 1}{(10.580,5.388)}
\gppoint{gp mark 1}{(10.618,6.684)}
\gppoint{gp mark 1}{(10.655,5.519)}
\gppoint{gp mark 1}{(10.693,5.645)}
\gppoint{gp mark 1}{(10.731,5.481)}
\gppoint{gp mark 1}{(10.768,5.102)}
\gppoint{gp mark 1}{(10.806,4.711)}
\gppoint{gp mark 1}{(10.844,5.645)}
\gppoint{gp mark 1}{(10.881,6.034)}
\gppoint{gp mark 1}{(10.919,6.194)}
\gppoint{gp mark 1}{(10.093,1.319)}
\gpcolor{color=gp lt color border}
\draw[gp path] (1.504,7.251)--(1.504,0.985)--(10.919,0.985)--(10.919,7.251)--cycle;
\gpdefrectangularnode{gp plot 1}{\pgfpoint{1.504cm}{0.985cm}}{\pgfpoint{10.919cm}{7.251cm}}
\end{tikzpicture}
}
    \subcaption{Program size controlled.}\label{fig:rand-limit}
  \end{minipage}
  \caption{Random walk in neutral landscape of ASM variations of
Insertion Sort.  \small A series of populations of 100 neutral
variants from 1 to 250 edits away from the original program.  At each
step on the X-axis, the mean number of lines of code and mutational
robustness of the members of the population are shown.  In Panel
\ref{fig:rand-no-limit} the size of neutral variants is allowed to vary,
while Panel \ref{fig:rand-limit} allows only variants that are less
than or equal to the length of the original program (in ASM LOC).  On
the 32-bit machine used for this experiment \texttt{insertion-sort}
compiles to 175 assembly LOC.  Both the average size and the
mutational robustness of mutant variants are shown on the Y-axis, the
X-axis shows the cumulative number of mutation operators applied.
\label{fig:rand-walk}}
\end{figure}
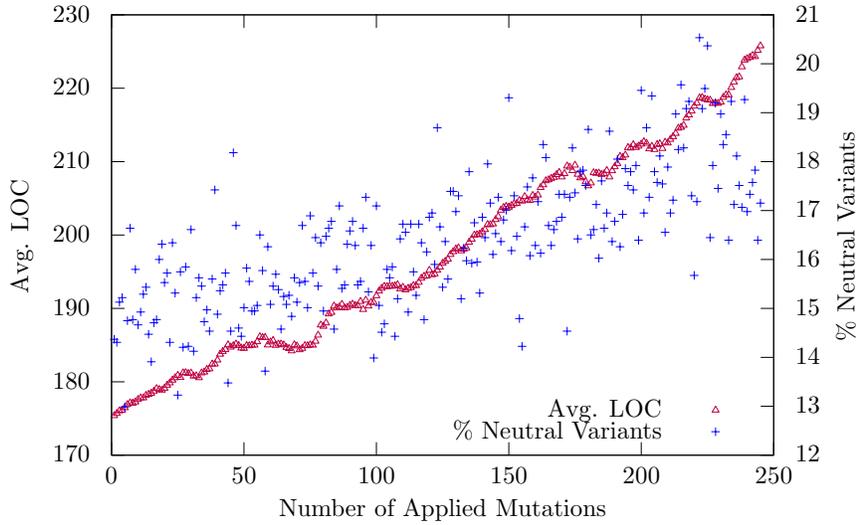
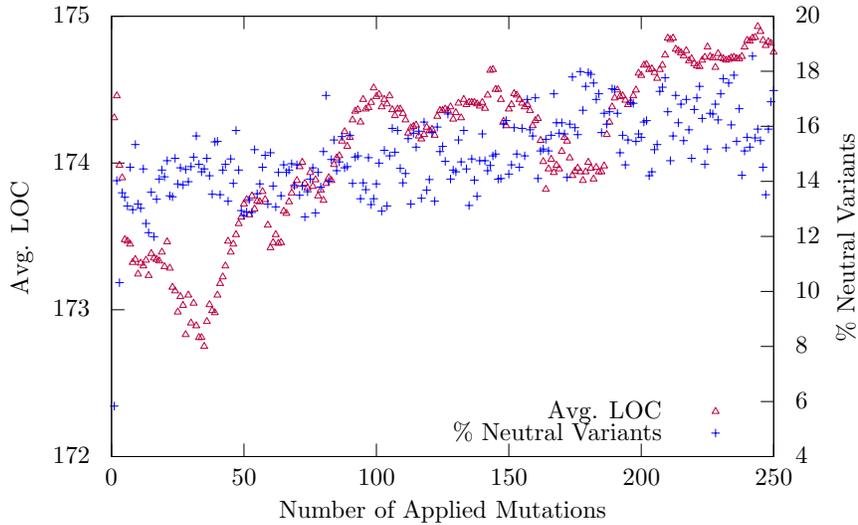

The previous experiments measured the percentage of first-order
mutations that are neutral. This subsection explores the effects of
accumulating cumulative neutral mutations in a small assembly program.
We begin with a working assembly code implementation of insertion
sort.  We apply random mutations using the ASM representation and
mutation operations defined in Section \ref{sec:rep-and-ops}.  After
each mutation, the resulting variant is retained if neutral, and
discarded otherwise.  The process continues until we have collected
100 neutral variants.  The mean program length and mutational
robustness of the individuals in this population are shown as the
leftmost red and blue points respectively in Figure
\ref{fig:rand-no-limit}.  From these 100 neutral variants, we then
generate a population of 100 second order neutral variants.  This is
accomplished by looping through the population of first-order mutants,
randomly mutating each individual once retaining the result if it is
neutral and discarding it otherwise.  Once 100 neutral second-order
variants have been accumulated, the procedure is iterated to produce
higher-order neutral variants.  This process produces neutral
populations separated from the original program by successively more
neutral mutations.  Figure \ref{fig:rand-walk} shows the results of
this process up to 250 steps producing a final population of 100
neutral variants, each of which is 250 neutral mutations away from the
original program.

The results show that under this procedure mutational robustness
increases with the mutational distance away from the original program.
This is not surprising given that at each step mutationally robust
variants are more likely to produce neutral mutants which will be
included in the subsequent population.  We conjecture that this result
corresponds to the population drifting away from the perimeter of the
program's neutral space.  Similar behavior has been described for
biological systems, where populations in a constant environment
experience a weak evolutionary pressure for increased mutational
robustness~\cite[Ch.~16]{wagner-rb}.

The average size of the program also increases with mutational
distance from the original program (Figure \ref{fig:rand-no-limit}),
suggesting that the program might be achieving robustness by adding
``bloat'' in the form of useless instructions.  To control for bloat,
Figure \ref{fig:rand-limit} shows the results of an experiment in
which only individuals that are the same size or smaller than
(measured in number of assembly instructions) the original program are
counted as neutral.  With this additional criterion, mutational
robustness continues to increase but the program size periodically
dips and rebounds, never exceeding the size of the original program.
The dips are likely consolidation events, where additional
instructions are discovered that can be eliminated.

This result shows that not only are there large neutral spaces
surrounding any given program implementation (in this instance,
permitting neutral variants as far as 250 edits removed from a
well-tested $<200$ LOC program), but they are easily traversable
through iterative mutation.  Figure \ref{fig:rand-limit} shows a small
increase in mutational robustness even when controlling for bloat.
Further experimentation will determine if these results generalize to
multiple programs.

\subsection{Multiple Languages}
\label{sec:multiple-languages}

\begin{table}
  \begin{center}
  \begin{tabular}{r|rrrr|rr}
           & C      & C++    & Haskell & OCaml  & Avg. & Std.Dev. \\
\toprule
 bubble    & 25.7   & 28.2   & 27.6    & 16.7   & 24.6 & 5.3      \\
 insertion & 26.0   & 42.0   & 35.6    & 23.7   & 31.8 & 8.5      \\
 merge     & 21.2   & 46.0   & 24.9    & 22.7   & 28.7 & 11.6     \\
 quick     & 25.5   & 42.0   & 26.3    & 11.4   & 26.3 & 12.5     \\
\midrule
 Avg.      & $24.6$ & $39.5$ & $28.6$  & $18.6$ & $27.9$          \\
 Std.Dev.  & $2.3$  & $7.8$  & $4.8$   & $5.7$  & $3.1$           \\
  \end{tabular}
  \end{center}
  \caption{\small Mutational robustness of sorting algorithms at the assembly
    instruction level with 100\%
    test suite coverage, for different algorithms and source
    language.
   \label{tab:mut-rb-over-langs}}
\end{table}

The source language and compilation process impose important
regularities on the assembly representations of programs.  In this
section we investigate how mutational robustness varies across
assembly code derived from different languages.  We evaluated the
mutational robustness of sorting algorithms compiled from five
languages that span three programming paradigms (imperative,
object-oriented, and functional).  We focused on sorting algorithms
because they are small and sufficiently well-understood to test
comprehensively: Test suites were hand-crafted to cover all executable
assembly instructions, branches, and corner cases.

The results shown in Table \ref{tab:mut-rb-over-langs} demonstrate
that mutational robustness is found across multiple programming
languages and paradigms. It is striking that even for a small,
comprehensively tested sorting program, on average 28\% of the
first-order mutations did not change the program's functionality.

This experiment addresses the question of whether our results depend
on idiosyncrasies of a particular language implementation or
programming paradigm, and they support the claim that mutational
robustness occurs at a significant rate in a wide variety of software.

\section{Application of Mutational Robustness to\\Proactive Bug Repair}
\label{sec:application}

The previous sections demonstrated the prevalence of mutational
robustness in software, independent of language, algorithm, or test
suite coverage. Section \ref{sec:cumulative} suggests that multiple
mutations can be accumulated in the same program without loss of
functionality.  This section provides one example of how we might
leverage software mutational robustness in a practical application
proactively repairing unknown bugs that are not covered by a program's
test suite.

We generate a population of multiple variations of a program, which
contain nontrivial changes to the algorithm or implementation but
remain within the program's neutral space.  Although these variants
are \emph{neutral} to the original program with respect to the
existing test suite, they may not be neutral to the original if the
test suite were later enhanced (e.g., by including tests for as yet
unknown bugs).  We find that some of these program variations are
immune to latent bugs in the original program, and they can be used to
automatically pinpoint repairs when new bugs arise.

The term \emph{artificial diversity} describes a wide variety of
techniques for automatically randomizing non-functional properties of
programs with the goal of disrupting widely replicated attacks.  Many
methods have been proposed, including stack frame layouts, instruction
set numberings, or address space layouts
~\cite{forrest-diversity,BarrantesEtAl05a,jackson2011compiler}.
Although the idea of diversifying certain aspects of a program's
implementation has been previously proposed~\cite{isr205}, neutral
mutations provide a much more general and practical approach.  We
extend earlier work in this area by generating program variants that
are distinct algorithmically but neutral with respect to the test
suite.  These distinct variants could be used in an \emph{n}-variant
system~\cite{nvariant}, in which a diverse population of programs is
run simultaneously on the same inputs.  When the neutral variants
contain algorithmic or implementation changes (rather than simple
remappings, as in address space layout randomization or instruction
set randomization), the approach is known as \emph{implementation
  diversity}~\cite{CowanEtAl00a}.  Such a system could be used to flag
potential bugs when there are discrepancies in observed behavior, and
may be more robust to bugs which only affect subsets of the
population.

This use of mutational robustness is analogous to software mutation
testing, with the critical differences that (1) neutral mutants are
retained rather than manually examined; (2) the test suite is not
augmented to kill all mutants; and (3) the set of mutation operators
considered is different.  The commercial practice of mutation testing
has been limited by the significant effort required to analyze mutants
that pass the test suite.  Such mutants must be manually classified,
either as fully equivalent to the original program or non-equivalent,
and the latter further classified as buggy or as superior to the
original program (cf. \emph{human oracle problem}
\cite{weyuker1982testing}).

Our proposed alternative to the traditional mutation testing practice
amortizes these labor intensive steps by retaining a population of all
such \emph{neutral} variants.  When a bug is encountered in the
original program, it will be detected if some members of the
population behave anomalously with respect to the result of the
population, and the non-failing variations need only then be analyzed
to suggest a repair. This approach of deferring analysis until a
potentially beneficial variation is found may be more feasible than
traditional mutation testing, because it does not require exhaustive
manual review of large numbers of program variants.

\subsection{Repairing Bugs}
\label{sec-diversity-bugs}

We first demonstrate that it is possible to construct variants in the
neutral fitness landscape that can repair unknown bugs while retaining
required functionality.  To do this, we seeded each of the programs in
Table \ref{tab:zak-total-bugs} with five random defects following an
established defect distribution~\cite[p.5]{fry10} and fault
taxonomy~\cite{knight85} (e.g., ``missing conditional clause,''
``extra statement,'' ``constant should have been variable,'' ``wrong
parameter''). The defects were seeded in advance and without regard to
the mutation operators. For each defect we produced a held-out test to
verify its presence or absence.

For each program, we generated 5,000 first-order neutral variants
using the mutation operators defined in Section \ref{sec:rep-and-ops}.
We then noted which of these passed any of the five held-out test
cases for the seeded bugs. In practice, 5,000 neutral variants proved
sufficient to generate at least one bug-repairing variant for most
programs, and if such a variant was generated at all, it was within
the first 5,000 neutral variants.  We tested this by searching up to
20,000 variants, which did not improve performance.  Only variants
that passed the original test suite were retained; the mutation
process did \emph{not} have access to the held-out test cases for the
seeded bugs.

Table \ref{tab:zak-total-bugs} shows the results. We say that a
variant repairs an unknown bug if it passed all original test cases
and the held-out test case associated with that bug.  We found repairs
which exactly revert the original bug as well as \emph{compensatory}
mutations which do not touch the bug itself but repair or avoid the
bug by changing other parts of the program.  Specifically, we found
that 3\% of the proactive repairs changed the same line of code in
which the original bug was seeded and 12\% of repairs affect code
within 5 lines of the seeded bug.  The remaining 88\% of repairs could
be considered compensatory in that they do not affect the bug directly
but rather change other portions of the program so that the bug is not
expressed.  In nature compensatory mutations are much more common than
mutations which directly repair a given fault~\cite{poon2005coupon}.

These proactive repairs can be used to \emph{pinpoint} the bug,
because the \lt{diff} between them and the original program locates
either the bug itself or relevant portions of the program code.
Previous work demonstrated that software engineers take less time to
address defect reports that are accompanied by such machine-generated
patch-like information~\cite{weimer06}, which provides evidence that
these proactive repairs would be useful in practice.

\begin{table}
\begin{center}
\begin{tabular}{lrr}
\toprule
Program              & Fraction of Bugs Fixed & Bug Fixes \\
\toprule
\texttt{bzip}        & 2/5                    & 63        \\
\texttt{imagemagick} & 2/5                    & 8         \\
\texttt{jansson}     & 2/5                    & 40        \\
\texttt{leukocyte}   & 1/5                    & 1         \\
\texttt{lighttpd}    & 1/5                    & 73        \\
\texttt{nullhttpd}   & 1/5                    & 7         \\
\texttt{oggenc}      & 0/5                    & 0         \\
\texttt{potion}      & 2/5                    & 14        \\
\texttt{redis}       & 0/5                    & 0         \\
\texttt{tiff}        & 0/5                    & 0         \\
\texttt{vyquon }     & 1/5                    & 1         \\
\bottomrule
average              & 1.0/5                  & 18.8      \\
\bottomrule
\end{tabular}
\end{center}
\caption{Bugs proactively repaired by neutral variants.  \small Five
unique bugs were seeded in each subject program according to a defect
distribution taken from the Firefox open source project. Five thousand
neutral variants were created for each program through AST level
mutation, without regard to the seeded bugs. Each variant passed all
visible tests. The ``Unique Bugs Fixed'' column counts the number of
seeded bugs fixed by at least one variant. The ``Bug Fixes'' column
counts the number of variants that fixed at least one bug.
\label{tab:zak-total-bugs} }
\end{table}

We observe some common trends when examining Table
\ref{tab:zak-total-bugs}.  The bugs repaired most easily were those
that naturally mirror the mutation strategies employed by our
technique.  For example, we found multiple examples of the repairs
that deleted problematic statements or clauses, corrected an incorrect
value for a constant, changed a relational operator (for instance
$\le$ to $<$), inserted clauses or statements to test for extra
conditions, changed a parameter value in a function call, etc.
However, there was significant overlap between the \emph{types} of
bugs that were proactively repaired and those that were not,
suggesting that additional tuning might improve results on these
currently unrepaired classes of bugs.  One bug that was never repaired
in our experiment involved an ``incorrect function call.''  To repair
this bug using our technique would require finding the ``correct''
function elsewhere in the program with exactly the correct parameters,
while avoiding extraneous mutations that change behavior on regression
test cases.  In this experiment, we focused on discovering the
proactive repair, and we leave for future work the problem of
automatically deciding how to resolve discrepancies that are
discovered among selected variants, either with an automated repair or
by generating additional test cases.

\begin{figure}[htb]
  \begin{center}
\adjustbox{width=1.0\textwidth}{
\begin{tikzpicture}[gnuplot]
\path (0.000,0.000) rectangle (12.500,8.750);
\gpcolor{color=gp lt color border}
\gpsetlinetype{gp lt border}
\gpsetlinewidth{1.00}
\draw[gp path] (1.136,0.985)--(1.316,0.985);
\node[gp node right] at (0.952,0.985) { 0};
\draw[gp path] (1.136,1.794)--(1.316,1.794);
\node[gp node right] at (0.952,1.794) { 1};
\draw[gp path] (1.136,2.603)--(1.316,2.603);
\node[gp node right] at (0.952,2.603) { 2};
\draw[gp path] (1.136,3.412)--(1.316,3.412);
\node[gp node right] at (0.952,3.412) { 3};
\draw[gp path] (1.136,4.221)--(1.316,4.221);
\node[gp node right] at (0.952,4.221) { 4};
\draw[gp path] (1.136,5.030)--(1.316,5.030);
\node[gp node right] at (0.952,5.030) { 5};
\draw[gp path] (1.136,5.839)--(1.316,5.839);
\node[gp node right] at (0.952,5.839) { 6};
\draw[gp path] (1.136,6.648)--(1.316,6.648);
\node[gp node right] at (0.952,6.648) { 7};
\draw[gp path] (1.136,7.457)--(1.316,7.457);
\node[gp node right] at (0.952,7.457) { 8};
\draw[gp path] (1.136,0.985)--(1.136,1.165);
\draw[gp path] (1.136,7.457)--(1.136,7.277);
\node[gp node center] at (1.136,0.677) { 0};
\draw[gp path] (2.265,0.985)--(2.265,1.165);
\draw[gp path] (2.265,7.457)--(2.265,7.277);
\node[gp node center] at (2.265,0.677) { 2};
\draw[gp path] (3.394,0.985)--(3.394,1.165);
\draw[gp path] (3.394,7.457)--(3.394,7.277);
\node[gp node center] at (3.394,0.677) { 4};
\draw[gp path] (4.523,0.985)--(4.523,1.165);
\draw[gp path] (4.523,7.457)--(4.523,7.277);
\node[gp node center] at (4.523,0.677) { 6};
\draw[gp path] (5.652,0.985)--(5.652,1.165);
\draw[gp path] (5.652,7.457)--(5.652,7.277);
\node[gp node center] at (5.652,0.677) { 8};
\draw[gp path] (6.780,0.985)--(6.780,1.165);
\draw[gp path] (6.780,7.457)--(6.780,7.277);
\node[gp node center] at (6.780,0.677) { 10};
\draw[gp path] (7.909,0.985)--(7.909,1.165);
\draw[gp path] (7.909,7.457)--(7.909,7.277);
\node[gp node center] at (7.909,0.677) { 12};
\draw[gp path] (9.038,0.985)--(9.038,1.165);
\draw[gp path] (9.038,7.457)--(9.038,7.277);
\node[gp node center] at (9.038,0.677) { 14};
\draw[gp path] (10.167,0.985)--(10.167,1.165);
\draw[gp path] (10.167,7.457)--(10.167,7.277);
\node[gp node center] at (10.167,0.677) { 16};
\draw[gp path] (10.167,0.985)--(9.987,0.985);
\node[gp node left] at (10.351,0.985) { 1};
\draw[gp path] (10.167,1.472)--(10.077,1.472);
\draw[gp path] (10.167,1.757)--(10.077,1.757);
\draw[gp path] (10.167,1.959)--(10.077,1.959);
\draw[gp path] (10.167,2.116)--(10.077,2.116);
\draw[gp path] (10.167,2.244)--(10.077,2.244);
\draw[gp path] (10.167,2.352)--(10.077,2.352);
\draw[gp path] (10.167,2.446)--(10.077,2.446);
\draw[gp path] (10.167,2.529)--(10.077,2.529);
\draw[gp path] (10.167,2.603)--(9.987,2.603);
\node[gp node left] at (10.351,2.603) { 10};
\draw[gp path] (10.167,3.090)--(10.077,3.090);
\draw[gp path] (10.167,3.375)--(10.077,3.375);
\draw[gp path] (10.167,3.577)--(10.077,3.577);
\draw[gp path] (10.167,3.734)--(10.077,3.734);
\draw[gp path] (10.167,3.862)--(10.077,3.862);
\draw[gp path] (10.167,3.970)--(10.077,3.970);
\draw[gp path] (10.167,4.064)--(10.077,4.064);
\draw[gp path] (10.167,4.147)--(10.077,4.147);
\draw[gp path] (10.167,4.221)--(9.987,4.221);
\node[gp node left] at (10.351,4.221) { 100};
\draw[gp path] (10.167,4.708)--(10.077,4.708);
\draw[gp path] (10.167,4.993)--(10.077,4.993);
\draw[gp path] (10.167,5.195)--(10.077,5.195);
\draw[gp path] (10.167,5.352)--(10.077,5.352);
\draw[gp path] (10.167,5.480)--(10.077,5.480);
\draw[gp path] (10.167,5.588)--(10.077,5.588);
\draw[gp path] (10.167,5.682)--(10.077,5.682);
\draw[gp path] (10.167,5.765)--(10.077,5.765);
\draw[gp path] (10.167,5.839)--(9.987,5.839);
\node[gp node left] at (10.351,5.839) { 1000};
\draw[gp path] (10.167,6.326)--(10.077,6.326);
\draw[gp path] (10.167,6.611)--(10.077,6.611);
\draw[gp path] (10.167,6.813)--(10.077,6.813);
\draw[gp path] (10.167,6.970)--(10.077,6.970);
\draw[gp path] (10.167,7.098)--(10.077,7.098);
\draw[gp path] (10.167,7.206)--(10.077,7.206);
\draw[gp path] (10.167,7.300)--(10.077,7.300);
\draw[gp path] (10.167,7.383)--(10.077,7.383);
\draw[gp path] (10.167,7.457)--(9.987,7.457);
\node[gp node left] at (10.351,7.457) { 10000};
\draw[gp path] (1.136,7.457)--(1.136,0.985)--(10.167,0.985)--(10.167,7.457)--cycle;
\gpcolor{color=black}
\node[gp node center,rotate=-270] at (0.246,4.221) {Number of Unique Bugs Fixed};
\gpcolor{color=black}
\node[gp node center,rotate=-270] at (11.792,4.221) {Number of Selected Variants};
\gpcolor{color=black}
\node[gp node center] at (5.651,0.215) {Number of Bugs Seeded};
\node[gp node right] at (7.493,8.108) {Number of Selected Variants};
\gpcolor{color=gp lt color 7}
\gpsetlinetype{gp lt plot 3}
\gpsetlinewidth{2.00}
\draw[gp path] (7.677,8.108)--(8.593,8.108);
\draw[gp path] (2.829,6.517)--(2.829,6.868);
\draw[gp path] (3.394,6.517)--(3.394,6.868);
\draw[gp path] (3.958,3.819)--(3.958,5.713);
\draw[gp path] (4.523,0.985)--(4.523,5.693);
\draw[gp path] (5.087,0.985)--(5.087,2.352);
\draw[gp path] (5.652,0.985)--(5.652,2.352);
\draw[gp path] (6.216,0.985)--(6.216,2.352);
\draw[gp path] (6.780,0.985)--(6.780,2.352);
\draw[gp path] (7.345,0.985)--(7.345,2.352);
\draw[gp path] (7.909,0.985)--(7.909,2.352);
\draw[gp path] (8.474,0.985)--(8.474,2.352);
\draw[gp path] (9.038,0.985)--(9.038,2.352);
\draw[gp path] (9.603,0.985)--(9.603,2.352);
\gpsetpointsize{4.00}
\gppoint{gp mark 4}{(2.829,6.715)}
\gppoint{gp mark 4}{(3.394,6.715)}
\gppoint{gp mark 4}{(3.958,5.272)}
\gppoint{gp mark 4}{(4.523,5.174)}
\gppoint{gp mark 4}{(5.087,1.959)}
\gppoint{gp mark 4}{(5.652,1.959)}
\gppoint{gp mark 4}{(6.216,1.959)}
\gppoint{gp mark 4}{(6.780,1.959)}
\gppoint{gp mark 4}{(7.345,1.959)}
\gppoint{gp mark 4}{(7.909,1.959)}
\gppoint{gp mark 4}{(8.474,1.959)}
\gppoint{gp mark 4}{(9.038,1.959)}
\gppoint{gp mark 4}{(9.603,1.959)}
\gppoint{gp mark 4}{(8.135,8.108)}
\gpcolor{color=gp lt color border}
\node[gp node right] at (7.493,8.416) {Number of Unique Bugs fixed};
\gpcolor{color=gp lt color 2}
\gpsetlinewidth{6.00}
\gppoint{gp mark 3}{(1.700,0.985)}
\gppoint{gp mark 3}{(2.265,0.985)}
\gppoint{gp mark 3}{(2.829,1.794)}
\gppoint{gp mark 3}{(3.394,1.794)}
\gppoint{gp mark 3}{(3.958,2.603)}
\gppoint{gp mark 3}{(4.523,3.412)}
\gppoint{gp mark 3}{(5.087,4.221)}
\gppoint{gp mark 3}{(5.652,5.030)}
\gppoint{gp mark 3}{(6.216,5.030)}
\gppoint{gp mark 3}{(6.780,5.030)}
\gppoint{gp mark 3}{(7.345,5.030)}
\gppoint{gp mark 3}{(7.909,5.030)}
\gppoint{gp mark 3}{(8.474,5.839)}
\gppoint{gp mark 3}{(9.038,5.839)}
\gppoint{gp mark 3}{(9.603,6.648)}
\gppoint{gp mark 3}{(8.135,8.416)}
\gpcolor{color=gp lt color border}
\gpsetlinetype{gp lt border}
\gpsetlinewidth{1.00}
\draw[gp path] (1.136,7.457)--(1.136,0.985)--(10.167,0.985)--(10.167,7.457)--cycle;
\gpdefrectangularnode{gp plot 1}{\pgfpoint{1.136cm}{0.985cm}}{\pgfpoint{10.167cm}{7.457cm}}
\end{tikzpicture}
}
    \caption{Number of unique bugs repaired and number of unique
      variants required to repair at least one seeded bug as a
      function of number of bugs seeded for the \texttt{potion}
      program.  \small Starting with a population of 5,000 neutral
      program variants the total number of unique bugs repaired is
      shown in blue and the number of neutral variants needed to find
      a repair for at least one seeded bug is shown in red.  The
      X-axis is the number of bugs seeded in the subject program
      (\texttt{potion}).  The Pearson correlation coefficient between
      the number bugs seeded and repaired is
      $0.95$.\label{fig:seeded-vs-fixed}}
  \end{center}
\end{figure}
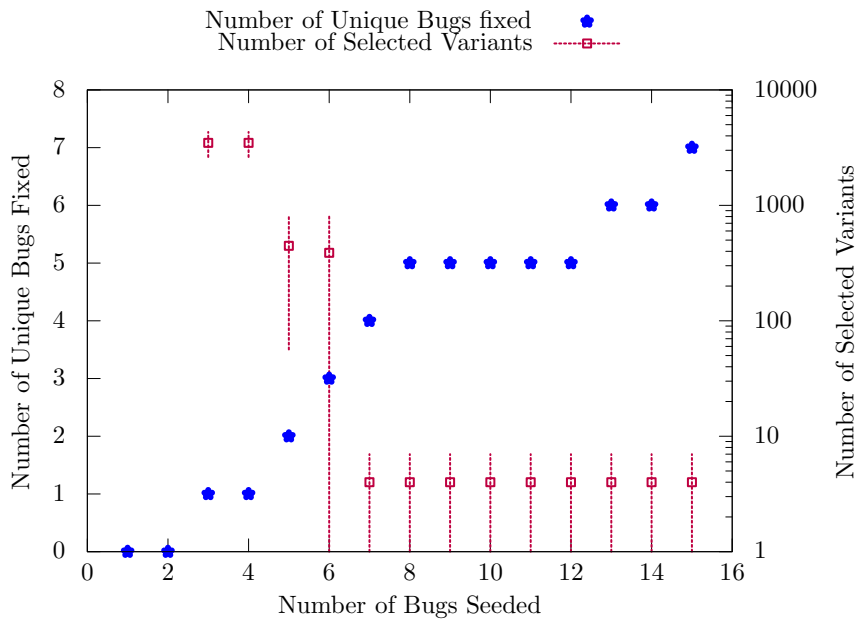

We predict that the more latent bugs there are in a program, the more
likely it is that at least one of them will be repaired through
proactive diversity.  This is relevant because most deployed programs
have significantly more than five outstanding defects (e.g., 18,165
from October 2001 to August 2005 and 2,013 open bug tickets May 2003
to August 2005 for Eclipse (V3.0) and Firefox (V1.0) respectively
\cite{anvik05}).  To test this prediction, we seeded one of our test
programs, \texttt{potion}, with ten additional held-out defects.  We
then generate 5,000 neutral variants which were selected to maximize
the number the distinct positions in the original program which they
modify.  Figure \ref{fig:seeded-vs-fixed} plots the number of distinct
bugs repaired by these 5,000 variants as a function of the number of
defects seeded, yielding a correlation of 95\%.  If the linear
relation shown in Figure \ref{fig:seeded-vs-fixed} applies to the
Eclipse and Firefox projects, a population of 5,000 program variants
could proactively repair as many as 9,000 and 1,000 latent bugs in
those systems, respectively.

We leave as future work the application of this technique to higher
order neutral mutants as constructed in Section \ref{sec:cumulative}.
While such variants would also greatly increase diversity they would
be less useful in directly pinpointing the source code implicated in
buggy behavior.

Although preliminary, these results show how the mutational robustness
properties of programs might be used to quickly repair programs or as
in the scenario outlined in the next subsection.

\subsection{Neutral variants for N-version programming}
\label{sec:neutral-n-version}

In their classic work on independence in multiversion programming,
Knight and Leveson found that distinct teams of humans, when given an
identical programming task, produce solutions that have correlated
errors (bugs)~\cite{knight-leveson}. That is, two independent teams
are unlikely to produce two independent implementations, which
decreases the potential benefit of N-person programming.

Previous work has shown that GP can serve as a promising tool for
N-version programming \cite{feldt1998generating}.  In particular, GP
can automate the significant task of developing N independent software
instances. However, the technique had limited applicability because it
relied on de-novo programs evolved in simplified languages
specifically designed for GP.

The diverse populations of program variants described in this section
could serve as the bases for an N-version programming system.  Such a
system could potentially address part of the non-independence hurdle
because the mutations are generated randomly rather than by people,
producing independent algorithmic changes. For example, if we seed 15
bugs in \texttt{potion} and then select at random ten neutral variants
which each proactively repair one bug, we see that, on average, 1.7
different defects are fixed by those ten proactive repairs, rather
than 1.0, which we would expect if they were 100\% correlated.

To put this idea into perspective, for the first month after Firefox
4.0 was released (March 11 through April 10, 2011), the project's
Bugzilla database shows that an average of 77 new, non-duplicate bugs
were reported per day.  Given that this rate of bug reporting is more
than ten times the number of bugs seeded in our evaluation, we
conjecture that an N-version programming system populated by our
technique would be at least as effective in similar real-world systems
as it is in our experimental setup.

\section{Discussion}
\label{sec:discussion}

The results presented here contradict the prevailing folk wisdom that
software is a precise and intentionally engineered mechanism, which is
brittle to small perturbations.  We find software to be inherently
robust to random mutations, malleable within extensive neutral
landscapes, and evolvable by combining the mutation operators
described here with crossover and selection
\cite{icse09,genprog-tse-journal,genprog-icse2012}.

This new view of software suggests a number of exciting areas for
future work.  These include further analysis of the extent, origins
and mechanics of software mutational robustness, novel practical
applications to software engineering, and new parallels between
software applications and evolved biological organisms.  We next
discuss threats to the validity of our findings, and then explore each
of these three areas of future work in turn.

\subsection{Threats to Validity}
\label{sec:threats-to-validity}

The quantitative results that we report depend on the particular
choice of mutation operators, and a competent programmer could likely
craft operators capable of achieving any pre-set desired level of
mutational robustness.  However, we believe that the operators used in
this paper are sufficiently simple, powerful and general that they
expose software mutational robustness as an inherent property of
software.  A topic left for future investigation is to define and test
a wider variety of mutation operators (including mutation operations
taken from the mutation testing community such as
Mothra~\cite{king1991fortran}), studying their effect on software
mutational robustness.

We use test suites to assess program behavior.  Insufficient test
suites could artificially inflate our estimates of software mutational
robustness.  Multiple aspects of our experimental design addresses
this threat explicitly.  In Section \ref{sec:benchmark} we selected
benchmarks programs with a wide variety of test-suite coverage and
depth, including both the Siemens programs, whose test suites have
been developed by multiple independent researchers to achieve an
exceptionally high quality, and small sorting programs, where we can
test all corner cases exhaustively.  In Section
\ref{sec:test-suite-quality} we analyze mutational robustness across
these three categories of programs, finding little variation in the
average and minimal mutational robustness measured for each
category. As a further step to address concerns about test-suite
quality, in Section \ref{sec:taxonomy} we manually categorized neutral
variants for a small program, finding that at least 29 of the 35
analyzed variants could not possibly fail any test suite that conforms
to the program specification.

In Section \ref{sec:application} we demonstrate applications of
software mutational robustness to the tasks of software development
and maintenance.  This demonstration does not address the impact which
low-quality test suites may have on the performance of our
demonstrated applications.  Such an investigation may be required
before such techniques are applied.

\subsection{Further Investigation}
\label{sec:more-soft-rb}

Although we demonstrate the reach of neutral software landscapes in
Section \ref{sec:cumulative} by evolving neutral variants hundreds of
edits removed from an original program, much about these landscapes
remains unknown.  How densely do the spaces of neutral variants fill
the space of all possible programs?  Are these neutral spaces
connected and traversable, or isolated?  Do these spaces extend to all
portions of program space (as the neutral spaces of mappings of RNA
sequences to RNA structures do \cite{schuster1994sequences})?

Measuring the ``distance'' between neutral program variants may
illuminate the effective impact of neutral program mutations.  Such
distance metrics could be based on information flow through the
program during execution~\cite{suh2004secure}, or comparison of stack
traces or system calls~\cite{forrestetal08a}.

Recent work has shown that Markov Chain Monte Carlo
\cite{andrieu2003introduction} (MCMC) techniques are capable of
traversing neutral program spaces defined by vastly smaller programs
on the order of 10 assembly instructions
\cite{schkufza2013stochastic}.  It remains unclear if these techniques
are extensible to the spaces defined by full-size programs with loops.

Software mutational robustness measures the density of neutral
variants within a single edit of a program.  However, the density of
higher-order neutral variants remains unknown.  We could possibly use
MCMC to sample the space of more neutral variants within a given edit
radius or the original program as follows:
\begin{enumerate}
\item Calculate the mixing time of an edit-radius constrained neutral
  variant markkov chain;
\item Record the neutrality of program encountered after the mixing time;
\item Use mark and recapture techniques from population ecology
  \cite{robson1964sample} to estimate the total volume of neutral
  variants.
\end{enumerate}
Performing mark and recapture experiments within neutral landscapes
found along the post mixing time walks would allow us to estimate the
absolute number of neutral variants within a given edit radius of the
original program.

This work does not address execution time software robustness.
Examples include: short error propagation distances observed in web
servers studied in \emph{failure oblivious}
computing~\cite{failure-oblivious}, and calculations such as those
prevalent in the PARSEC benchmark suite~\cite{bienia2008parsec}, which
iteratively converge on the correct solutions and which may become
more common as future applications leverage multi-core architectures.

\subsection{Applications to Software Engineering}
\label{sec:software-engineering-applications}

There are an infinite number of possible implementations for any
functional program specification, and Table \ref{tab:benchmarks-rb}
can be interpreted as illustrating how easy it is to find such
multiple equivalent implementations.  These results also suggest that
programs have a significant amount of unidentified and unexploited
redundancy.  One practical implication of large traversable neutral
landscapes would be to search the neutral landscapes for regions that
optimize non-functional properties such as program size (as
demonstrated in Figure \ref{fig:rand-limit}), memory requirements,
runtime, power consumption, or any other non-functional software
features.  This may explain the success of previous work optimizing
graphics shader software~\cite{genprog-shader}.

The large neutral landscapes revealed by our study may also explain
the success of evolutionary methods, such as GP, on the task of
automated program repair~\cite{icse09}.  Although this parallel with
evolutionary biology is intriguing (Section \ref{sec:background-bio}),
we do not yet have definitive evidence that quantifies the role of
mutational robustness in automated program repair, a topic that we
leave for future work.  We suspect that other methods of automated
repair
(e.g.,~\cite{dallmeier-ase-2009,liblit2011,zeller2010,icse09,rinardClearview})
may ultimately be understood in the context of software mutational
robustness and evolvability.

In addition to the proactive diversity application described in
Section \ref{sec:application}, it might be feasible to incorporate
other machine learning methods into software development and
maintenance processes; such methods typically work poorly on brittle
system and require some degree of robustness while they search for
improved solutions.

Our results constitute a fundamentally different interpretation of
software mutants than that of the mutation testing community.  We do
not view software mutants as faulty or neutral mutants as an indicator
of an insufficient test suite.  Instead we view software mutants as
natural and neutral mutants as alternate implementations of software
specifications which admit many non-equivalent implementations.  The
quick-sort example discussed in Section \ref{sec:intro} demonstrates
that simple mutations can yield different fully correct
implementations. Yin \emph{et al.}  provide another example in which a
fully-formally verified implementation of AES encryption is changed
based on what was ``purely an implementation decision, [where] the
specification did not impose any restrictions,'' yielding another
formally verified implementation~\cite[p.61]{yin09}.

Such alternate implementations, which may have distinct runtime
properties but still conform to the program specification, are
\emph{neutral}, but are not \emph{equivalent} in the sense used by the
mutation testing community.  The authors have performed a thorough
review of the mutation testing literature to compare our empirical
findings of software mutational robustness to typical frequencies of
equivalent mutants in the mutation testing paradigm.  The mutation
testing community does identify equivalent mutants as one of the
fundamental problems in mutation testing (cf. \cite[Section
II.C]{mutation-testing}).  Although this problem is well known, we
were unable to find formal publications that experimentally identify
the fraction of equivalent mutants (aside from work explicitly
targeting Object-Oriented mutation operators which generate
particularly high rates of equivalent mutants
\cite{segura2011mutation,schuler2010covering,offutt2006class}).

Through our own review of the mutation testing literature we collected
unreported counts of equivalent mutants from a number of papers
\cite{frankl1997all,offutt1996experimental,offutt1994using,demillo1991constraint}
that all used the Mothra \cite{king1991fortran} mutation operators and
that found equivalent mutant rates of 9.92\%, 6.75\%, 6.24\% and
6.17\% respectively.  Although these works used different mutation
operators than those described in Section \ref{sec:rep-and-ops}, the
percentages of equivalent mutants found are smaller than the
percentage of neutral mutants which we report, and are thus consistent
with our results because equivalent mutants are a strict subset of
neutral mutants.

Our re-interpretation of software mutants opens the possibility of
re-purposing the many tools developed by the mutation testing
community. For example, runtime optimization techniques such as
\emph{mutant scheme generation}~\cite{untch1995schema} enable the
compilation of ``super mutants'' capable of executing all first-order
mutants of a program.  Such a system could potentially be applied to
efficiently deploy and run populations of diverse software variants as
described in Section \ref{sec:application} on end user systems.
Techniques for the automatically identifying fully equivalent mutants
\cite{offutt1994using,offutt1997automatically} could be used to
differentiate between those neutral mutants that are identical to the
original program and those that encode a distinct implementation.
These are just two examples of how the extensive toolbox of mutation
testing could be re-purposed to leverage the view of software mutants
as neutral.

\subsection{Comparison to Biological Mutational Robustness}
\label{sec:comparison-to-biology}

Beyond the significance for computation, we believe that our results
are relevant to biologists.  We considered the effect of repeated
neutral mutations in a single program, showing that robustness can
increase systematically through population exploration of the neutral
landscape; a property shared with biological systems
\cite{wagner2008robustness}.  We also showed that it is possible to
generate programs that are many mutational steps removed from the
original while retaining functionality (i.e., without leaving the
neutral plateau).  These large extended neutral landscapes are thought
to be essential to the ability of biological systems to improve
through natural selection \cite{schuster1994sequences,van1999neutral}
but difficult to measure experimentally.  Our software analogs may
eventually provide a useful experimental framework for testing
hypotheses about the role of neutrality in biological evolution.

Currently, it is difficult to draw conclusions from a quantitative
comparison of the $36.8\%$ mutational robustness found in software to
typical levels of mutational robustness in biological systems (e.g,
$30\%$ mutational robustness in hominids \cite{eyre2007distribution},
or the almost $40\%$ neutrality of gene knockouts (deletions) in yeast
\cite{smith1996functional}).  There are many drivers of mutational
robustness in biological systems; environmental stability influences
levels of robustness \cite{meyers2005potential}, the centrality of a
gene may influence its mutational robustness
\cite{kafri2008preferential}, evolution may either select for
mutational robustness \cite{van1999neutral} or it may be more
effective in organisms that are mutationally robust
\cite{ciliberti2007innovation}.  There are analogs to each of these
factors in software systems, which future work may relate to the
levels of mutational robustness found in software.

Mutational robustness has many correlates in biological systems.
These include a correlation between environmental and mutational
robustness~\cite{lenski2006balancing,kitano2004biological,van1999neutral}
and a correlation between mutational robustness and
evolvability~\cite{wagner2008robustness}.  The presence of analogous
correlations in software systems is an intriguing possibility that
could be investigated empirically.  Such an investigation could
indicate whether these relations are general across complex
mutationally robust systems or are specific to biological systems.  If
such correlations do exist in software, they could lead to new
applications, such as methods of automated hardening which
automatically increase environmental robustness through increasing
mutational robustness (as in Section \ref{sec:cumulative}).

We may use the ratio between genome size and gene number as a proxy
for the relative mutational robustness of biological organisms,
although this would certainly be a lower bound.  However this ratio
varies widely from 97:47\% in prokaryotes and viruses, to 87:1\% in
eukaryotes.  We note that those portions of software that are executed
by the test suite (e.g., where we limit our mutation operations) could
be compared to those portions of the genome that code for genes.  In
both cases we can be sure that those portions have a phenotypic
effect, and in both cases we can't say for sure that the uncovered or
non-coding portions have no effect.  Such portions may either regulate
expression or affect compilation respectively.  Direct software
analogs of non-coding regulatory DNA could include type annotations or
pragmas.

Similarly, some of the causes of large genomes in biological organisms
have immediate analogs in the would of software development.  Gene
duplication is thought to be a significant contributor to the growth
in genome size.  The analogous practice of copy-paste programming is a
wide spread and common software development technique
\cite{kim2004ethnographic,kamiya02,krinke07,juergens09}.

Ultimately software may stand with biological organisms as a second
example of an evolved system.  Albeit one in which humans engineers
are the mechanisms of both mutation and selection~\cite{ackley-talk}.

\section{Conclusion}
\label{sec:conclusion}

The previous sections described experimental results, using three
simple mutation operators, which show that software is surprisingly
robust to random mutations.  For the programs we tested, $37\%$ of the
mutations had no effect on software functionality, as measured by the
programs test suites.  Software mutational robustness, or neutrality,
is observed even in programs that are is completely correct according
to their specifications.  Just as neutrality is believed to enhance
evolvability in naturally evolving populations, so may software
neutrality enable and explain the evolvability of software, either
through automated means (e.g., \cite{icse09}) or by humans.

Software robustness is potentially useful for enhancing the resilience
of software systems.  We demonstrate this idea by describing a method
that increases software diversity, automatically generating software
variants that are immune to as yet undiscovered bugs.  The insights
into software described here suggest several opportunities for the
Software Engineering community, including the following: Creating
system diversity, for example, to protect against security exploits;
incorporating machine learning methods into software development and
maintenance; improving program performance; or developing
error-tolerant computations.

We postulate that the presence of mutational robustness in software is
not an effect of intentional design, but is rather an effect of
software's provenance through natural selection---even though the
agents of selection, mutation and reproduction are human engineers.
In this way, mutational robustness can be viewed as a property arising
through inadvertent selection in both natural and engineered systems.
Further study of software as an evolved system may yield new insights
into those aspects of evolution that are specific to biological
systems and those which are general across other complex evolved, and
even engineered, systems.  Because software is fundamentally easier to
instrument and observe than naturally occurring populations, studying
software robustness may lead in the future to an increased
understanding of the role of neutrality in natural evolution.

\section{Acknowledgments}
\label{sec:acknowledgments}
The authors would like to thank Lauren Ancel Meyers, William
B. Langdon and Peter Schuster for their thoughtful comments.  This
work was supported by the National Science Foundation (SHF-0905236),
Air Force Office of Scientific Research (FA9550-07-1-0532,
FA9550-10-1-0277), DARPA (P-1070-113237), DOE (DE-AC02-05CH11231) and
the Santa Fe Institute.

\end{document}